\documentclass[12pt, final]{article}

\usepackage{virgil_env} 
\usepackage{multirow}
\usepackage{fullpage}
\usepackage{nips10submit_e}
\nipsfinalcopy
\usepackage{nips10submit_e}
\nipsfinalcopy

\definecolor{mygray}{RGB}{153,153,153}
\definecolor{myred}{RGB}{221,77,57}
\definecolor{myblue}{RGB}{68,104,252}
\definecolor{mygreen}{RGB}{86,149,22}

\renewcommand*{\P}{\ensuremath{\mathbf{P}}\xspace}

\newcommand*{\bin}[1]{\texttt{#1}}

\newcommand*{\setX}{\ensuremath{\mathbf{X}}\xspace}

\newcommand*{\ei}{\ensuremath{{\tt ei}}\xspace}

\newcommand*{\Probstar}[1]{ \ensuremath{ \textstyle{\Pr^*} \!\! \left( #1 \right)}}

\usepackage{authblk}
\author{Virgil Griffith}
\affil{Computation and Neural Systems, Caltech, Pasadena, CA 91125 \\ \texttt{i@virgil.gr}}

\usepackage{wasysym}

\usepackage{mathenv}
\usepackage{wasysym}

\title{A Principled Infotheoretic $\phi$-like Measure}

\newtheorem{lem}{Lemma}

\newcommand*{\Icupe}[2]{\ensuremath{\opname{I}_{\cup}\!\left( #1 \! : \! #2 \right)}}

\newcommand{\IbP}{\textnormal{IcP}}
\newcommand{\IbPTwo}{\textnormal{IcB}}

\newcommand*{\setP}{\ensuremath{\mathbf{P}}}

\newcommand{\Szero}{$\mathbf{\left(S_0\right)}$\xspace}

\newcommand{\Mone}{$\mathbf{\left(M_1\right)}$\xspace}

\renewcommand*{\wedge}{\curlywedge}

\newcommand*{\bpsi}{\ensuremath{\langle \psi \rangle}}

\begin{document}

\maketitle

\vspace{-0.05in}
\begin{abstract}
Integrated information theory \cite{balduzzi-tononi-08,balduzzi09,tononi-08} is a mathematical, quantifiable theory of conscious experience.  The linchpin of this theory, the $\phi$ measure, quantifies a system's irreducibility to disjoint parts.  Purely as a measure of irreducibility, we pinpoint three concerns about $\phi$ and propose a revised measure, $\psi$, which addresses them.  Our measure $\psi$ is rigorously grounded in Partial Information Decomposition and is faster to compute than $\phi$.
\end{abstract}

\section{Introduction}

The measure of integrated information, $\phi$, is an attempt to a quantify a neural network's magnitude of conscious experience.  It has a long history \cite{tononi04, balduzzi-tononi-08, tononi12}, and at least three different measures have been called $\phi$.  Conceptually, the $\phi$ measure aims to quantify a system's ``functional irreducibility to disjoint parts''.  Although innovative, the $\phi$ measure from \cite{balduzzi-tononi-08} has some peculiarities.  Using Partial Information Decomposition (PID), we derive a principled info-theoretic measure of irreducibility to disjoint parts\cite{qirred}; our PID-derived measure, $\psi$, has numerous desirable properties over the $\phi$ from \cite{balduzzi-tononi-08}.

We aim for $\psi$ to be a principled, well-behaved $\phi$-like measure that resides purely within Shannon information theory.  We compare $\psi$ to the older $\phi$ measure from \cite{balduzzi-tononi-08} because it is the most recent purely information-theoretic $\phi$.  We recognize that the most recent version of $\phi$\cite{tononi12} knowingly and purposely sits outside standard information theory.\footnote{The most recent version of $\phi$ \cite{tononi12} utilizes the Earth Mover's Distance among states and thus varies with the chosen labels of the states.  Although less of an issue for binary systems, a canonical property of information theories spanning from Shannon to Kolmogorov (algorithmic information theory) is invariance under relabeling of states.}$^,$\footnote{If one wished to use $\psi$ within the larger ``big phi'' conceptual framework per \cite{tononi12} you would replace all instances of the measure ``small phi'' with $\psi$.}

\section{Preliminaries}

We use the following notation throughout.

\begin{description}
	\item[$n$:] the number of indivisible elements in network $X$. $n \geq 2$.
	\item[$\mathbf{P}$:] a partition of the $n$ indivisible nodes clustered into $m$ parts.  Each part has at least one node and each partition has at least two parts, so $2 \leq m \leq n$.
	\item[$X^\P_i$:] a random variable representing a part $i$ at time=$0$.  $1 \leq i \leq m$.
	\item[$Y^\P_i$:] a random variable representing part $i$ after $t$ updates.  $1 \leq i \leq m$.
	\item[$X$:] a random variable representing the entire network at time=$0$.  $X \equiv X^\P_1 \cdots X^\P_m$.
	\item[$Y$:] a random variable representing the entire network after $t$ applications of the neural network's update rule.  $Y \equiv Y^\P_1 \cdots Y^\P_m$.
	\item[$y$:] a single state of the random variable $Y$.
	\item[$\setX$:] The set of the $n$ indivisible elements at time=$0$.
\end{description}

For readers accustom to the notation in \cite{balduzzi-tononi-08} the translation is: $X \equiv X_0$, $\ Y \equiv X_1$, $\ X^\P_i \equiv M_0^i$, and $Y^\P_i \equiv M_1^i$.

For pedagogical purposes we confine this paper to deterministic neural networks.  Therefore all remaining entropy at time $t$ conveys information about the past, i.e., $\info{X}{Y} = \ent{Y}$ and $\info{X}{Y^\P_i} = \ent{Y^\P_i}$ where $\info{\bullet}{\bullet}$ is the mutual information and $\ent{\bullet}$ is the Shannon entropy\cite{cover-thomas-91}.  Our model generalizes to probabilistic units with any finite number of discrete---but not continuous---states\cite{seth-2010}.  All calculations are in \emph{bits}.

\subsection{Model Assumptions}
\label{assumptions}
\begin{enumerate}
    \item[\textbf{(A)}] The $\phi$ measure is a \emph{state-dependent} measure.  Meaning that every output state $y \in Y$ has its own $\phi$ value.  To simplify cross-system comparisons, some researchers\cite{seth-2010} prefer to consider only the averaged $\phi$, denoted $\bphi$.  Here we adhere to the original theoretical state-dependent formulation.  However, when comparing large numbers of networks we use $\bphi$ for convenience.

	\item[\textbf{(B)}] 	The $\phi$ measure aims to quantify ``information intrinsic to the system''.  This is often thought to be synonymous with causation, but it's not entirely clear.  But for this reason, in \cite{balduzzi-tononi-08} all random variables at time=$0$, i.e., $X$ and $X^\P_1, \ldots, X^\P_m$ are forced to follow an \emph{independent discrete uniform distribution}.  There are actually several plausible choices for the distribution on $X$ (see Appendix \ref{appendix:Xdistribution}).  But for easier comparison to \cite{balduzzi-tononi-08}, here we also take $X$ to be an independent discrete uniform distribution.  This means that $\forall i \not= j \ \ \info{X^\P_i}{X^\P_j}=0$ and $\ent{X} = \log_2 \left| X \right|$, $\ent{X^\P_i} = \log_2 \left| X^\P_i \right|$ where $|\bullet|$ is the number of states in the random variable.
    
    \item[\textbf{(C)}] We set $t=1$, meaning we compute these informational measures for a system undergoing a single update from time=$0$ to time=$1$.  This has no impact on generality (see Appendix \ref{appendix:singletime}).  To analyze real biological networks one would sweep $t$ over all reasonable timescales choosing the $t$ that maximizes the complexity metric.
    
\end{enumerate}

\section{How $\phi$ Works}
The $\phi$ measure has four steps and proceeds as follows.

\begin{enumerate}
\item For a given state $y \in Y$, \cite{balduzzi-tononi-08} first defines the state's \emph{effective information} quantifying the total magnitude of information the state $y$ conveys about $X$, the r.v. representing a maximally ignorant past.  This turns out to be identical to \cite{meister-99}'s ``specific-surprise'', $\info{X}{y}$,
\begin{equation}
    \ei(X \to y) = \info{X}{y} = \DKL{ \Prob{X|y} }{ \Prob{X}} \; .
\label{eq:EIspecificsurprise}    
\end{equation}

Given $X$ follows a discrete uniform distribution (assumption \textbf{(B)}), $\ei(X \to y)$ simplifies to,
\begin{equation}
\begin{split}
    \ei(X \to y) &= \ent{X} - \ent{X|y} \\
    &= \ent{X} - \sum_{x \in X} \Prob{x|y} \log_2 \frac{1}{\Prob{x|y}} \; ;
\end{split}
\label{eq:balduzziEI}
\end{equation}

in the nomenclature of \cite{nyberg09}, $\ei(X \to y)$ can be understood as the ``total causal power'' the system exerts when transitioning into state $y$.

\item The second step is to quantify how much of the total causal power isn't accounted for by the disjoint parts (partition) $\mathbf{P}$.  To do this, they define the \emph{effective information beyond partition $\boldsymbol{\mathit{P}}$},
\begin{equation}
    \ei\left(X \to y \middle/ \mathbf{P} \right) \equiv \DKL{ \Prob{X\middle|y} }{ \prod_{i=1}^m \Prob{X^\P_i \middle| y^\P_i} } \; .
\label{eq:eibeyondP}    
\end{equation}

The intuition behind $\ei(X \to y / \mathbf{P})$ is to quantify the amount of causal power in $\ei(X~\to~y)$ that is irreducible to the parts $\mathbf{P}$ operating independently.\footnote{In \cite{balduzzi-tononi-08} they deviated slightly from this formulation using a process termed ``perturbing the wires''.  However, subsequent work\cite{tononi-08,tononi12} disavowed perturbing the wires and thus we don't use it here.  For discussion see Appendix \ref{appendix:perturbingwires}.}

\item After defining the causal power beyond an arbitrary partition $\mathbf{P}$, the third step is to find the partition that accounts for as much causal power as possible.  This partition is called the \emph{Minimum Information Partition}, or MIP.  They define the MIP for a given state $y$ as,\footnote{In \cite{balduzzi-tononi-08} they additionally consider the \emph{total partition} as a special case, meaning $m=1$ and $X_1^\P = X$.  However, subsequent work\cite{tononi-08,tononi12} disavowed the total partition and thus we don't use it here.}
\begin{equation}
    \textnormal{MIP}(y) \equiv \underset{\mathbf{P}}{\argmin} \ \frac{ \ei( X \to y / \mathbf{P} ) }{ (m - 1) \cdot \min_i \ent{X^\P_i} } \; .
\label{eq:phiMIP}
\end{equation}

Finding the MIP of a system by brute force is incredibly computationally expensive---enumerating all partitions of $n$ nodes scales $O(n!)$ and even for supercomputers becomes intractable for $n > 32$ nodes.

\item Fourth and finally, the system's causal irreducibility (to disjoint parts) when transitioning into state $y \in Y$, $\phi(y)$, is the effective information beyond $y$'s MIP,
\begin{equation}
    \phi(y) \equiv \ei\left( X \to y \middle/ \mathbf{P}=\textnormal{MIP}(y) \right) \; .
\label{eq:phiphi}    
\end{equation}
\end{enumerate}

\subsection{Stateless $\phi$ is $\bphi$}
\label{section:bracketphi}
Per eq.~\eqref{eq:phiphi} $\phi$ is defined for every state $y \in Y$, and a single system can have wide range of $\phi$-values.  In \cite{seth-2010}, they found this medley of state-dependent $\phi$-values unwieldy, and wanted a single number for each system.  They achieved this by averaging the effective information over all states $y$.  This results in the four corresponding stateless measures:
\begin{equation}
\begin{split}
    \left\langle \ei(Y) \right\rangle &\equiv \mathbb{E}_{y} \ei(X \to y) = \info{X}{Y} \\
    \left\langle \ei(X \to Y/\mathbf{P}) \right\rangle &\equiv \mathbb{E}_{y} \ei\left(X \to y \middle/ \mathbf{P} \right) = \info{X}{Y} - \sum_{i=1}^m \info{X^\P_i}{Y^\P_i} \\
    \left\langle \textnormal{MIP} \right\rangle &\equiv \underset{\mathbf{P}}{\argmin} \frac{ \left\langle \ei(Y/\mathbf{P}) \right\rangle}{(m - 1) \cdot \min_i \ent{X^\P_i}} \\
    \bphi &\equiv \left\langle \ei\left( Y  \middle/ \mathbf{P}=\langle\textnormal{MIP}\rangle \right) \right\rangle \; .
\end{split}
\label{eq:bracketphi}
\end{equation}

Although the distinction has yet to affect qualitative results, researchers should note that $\bphi \not= \mathbb{E}_{y} \phi(y)$.  This is because whereas each $y$ state can have a different MIP, for $\bphi$ there's only one MIP for all states.

\section{Three Concerns about $\phi$}
\label{sect:oddities}
\begin{figure}[h!]
\begin{minipage}{0.43\textwidth}	
	\raggedleft
	\subfloat[OR-GET network]{ \includegraphics[width=1.5in]{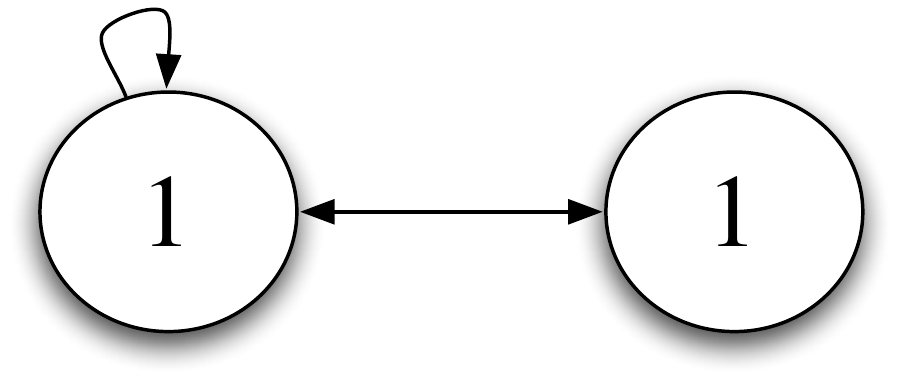} \label{fig:OG} }
	
	\subfloat[OR-XOR network]{ \includegraphics[width=1.5in]{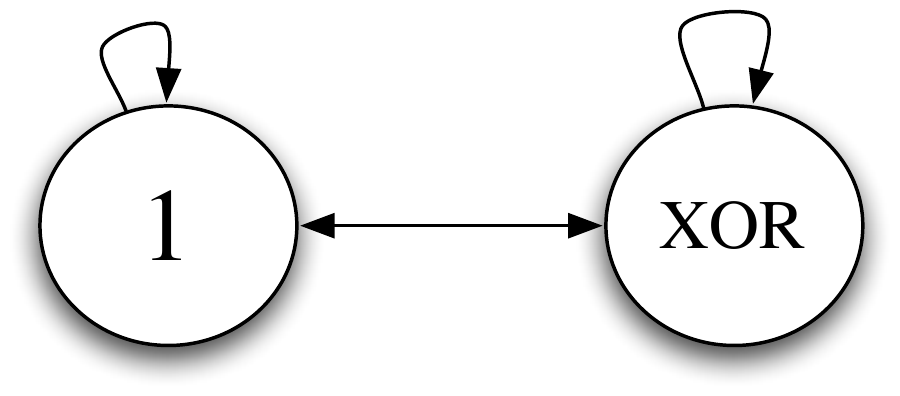} \label{fig:OX} }	
\end{minipage} \begin{minipage}{0.45\textwidth}
	\centering
	\begin{tabular}{c c c c} \toprule
		\multirow{2}{*}{$X$} & \ & OR- & OR-\\
		\ &  \ & GET & XOR \\
	\midrule
		\bin{00} & $\to$ & \bin{00} & \bin{00} \\
		\bin{01} & $\to$ & \bin{10} & \bin{11} \\
		\bin{10} & $\to$ & \bin{11} & \bin{11} \\
		\bin{11} & $\to$ & \bin{11} & \bin{10} \\
	\bottomrule
	\end{tabular}
		\caption*{Transition table for (a), (b)}
\end{minipage}

\vspace{0.25in}
\centering

\begin{tabular}{  l c c c c | c c c c } \toprule
\ & \multicolumn{4}{c}{OR-GET \subref{fig:OG} } & \multicolumn{4}{c}{OR-XOR \subref{fig:OX}} \\
\midrule
	& \bin{00} & \bin{01} & \bin{10} & \bin{11} & \bin{00} & \bin{01} & \bin{10} & \bin{11} \\
\midrule
    $\Prob{y}$  & $\nicefrac{1}{4}$ & - & $\nicefrac{1}{4}$ & $\nicefrac{1}{2}$ & $\nicefrac{1}{4}$ & - & $\nicefrac{1}{4}$ & $\nicefrac{1}{2}$ \\
	$\ei(y)$  & 2.00 & - & 2.00 & 1.00          & 2.00 & - & 2.00 & 1.00 \\
	$\phi(y)$ & 1.00 & - & \textbf{2.58} & 0.58 & 1.00 & - & 1.58 & \textbf{1.08} \\
\bottomrule
\end{tabular}

\caption{Example OR-GET shows that \ensuremath{\phi(y)} can exceed not only \ensuremath{\ei(X~\to~y)}, but $\ent{X}$!  A dash means that particular $y$ is unreachable for the network. The concerning $\phi$ values are \textbf{bolded}.}

\label{fig:exceeds}
\end{figure}

\textbf{$\phi(y)$ can exceed $\ent{X}$.}  \figref{fig:exceeds} shows examples OR-GET and OR-XOR.  On average, each looks fine---they each have $\ent{X}=2$, $\info{X}{Y}=1.5$, and $\bphi=1.189$  bits---nothing peculiar.  This changes when examining the individual states $y \in Y$.

For OR-GET, the $\phi(y=\bin{10})\approx 2.58$ bits.  Here $\phi(y)$ \emph{exceeds} the entropy of the entire system, $\ent{X Y} = \ent{X} = 2$ bits.  This means that for $y=\bin{10}$, the ``irreducible causal power'' exceeds not just the total causal power, $\ei(X \to y)$, but $\ei$'s upperbound of $\ent{X}$!  This is concerning.

For OR-XOR, $\phi(y=\bin{11}) \approx 1.08$ bits.  This does not exceed $\ent{X}$, but it does exceed the specific surprise, $\info{X}{y=\bin{11}} = 1$ bit.  Per eq.~\eqref{eq:bracketphi}, in expectation $\left\langle\ei(X \to Y/\mathbf{P}) \right\rangle \leq \info{X}{Y}$ for any partition $\mathbf{P}$.  The analogous information-theoretic interpretation for a single state would be more natural if likewise $\ei(X\!\to\!y/\mathbf{P}) \leq \info{X}{y}$ for any partition $\mathbf{P}$.

It's important to note neither issue is due to normalizing in eq.~\eqref{eq:phiMIP}.  For OR-GET and OR-XOR there's only one possible partition, and thus the normalization has no effect.  These oddities arise from the expression for the effective information beyond a partition, eq.~\eqref{eq:eibeyondP}.

\begin{figure}[h!]
\centering
\begin{minipage}{0.36\textwidth}	
	\raggedleft
	\subfloat[AND-ZERO network]{ \includegraphics[width=1.5in]{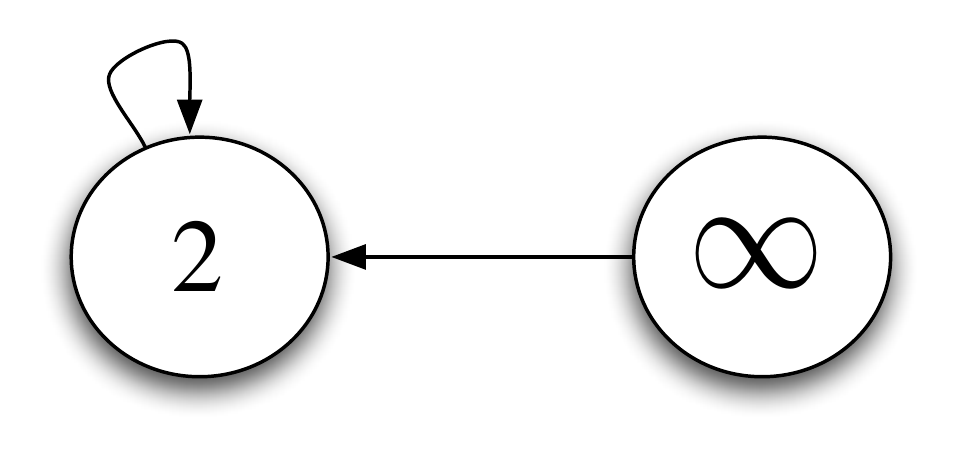} \label{fig:AZ} }
	
	\subfloat[AND-AND network]{ \includegraphics[width=1.5in]{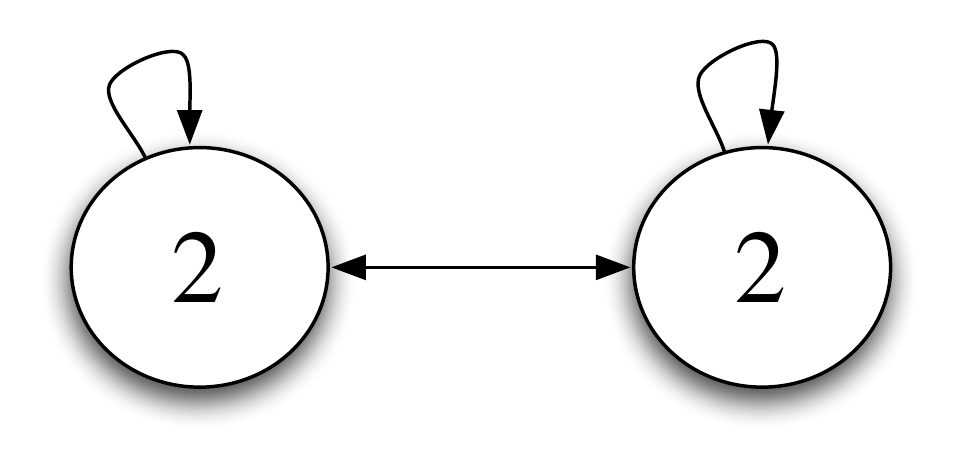} \label{fig:AA}  }
\end{minipage} \begin{minipage}{0.4\textwidth}
	\centering

	\begin{tabular}{c c c c} \toprule
		\multirow{2}{*}{$X$} & \ & AND- & AND-\\
		\ &  \ & ZERO & AND \\
	\midrule
		\bin{00} & $\to$ & \bin{00} & \bin{00} \\
		\bin{01} & $\to$ & \bin{00} & \bin{00} \\
		\bin{10} & $\to$ & \bin{00} & \bin{00} \\
		\bin{11} & $\to$ & \bin{10} & \bin{11} \\
	\bottomrule
	\end{tabular}
		\caption*{Transition table for (a), (b)}
\end{minipage}

\vspace{0.25in}

\begin{tabular}{  l c c c c | c c c c } \toprule
\ & \multicolumn{4}{c}{AND-ZERO \subref{fig:AZ} } & \multicolumn{4}{c}{AND-AND \subref{fig:AA}} \\
\midrule
	& \bin{00} & \bin{01} & \bin{10} & \bin{11} & \bin{00} & \bin{01} & \bin{10} & \bin{11} \\
\midrule
    $\Prob{y}$  & $\nicefrac{3}{4}$ & - & $\nicefrac{1}{4}$ & - & $\nicefrac{3}{4}$ & - & - & $\nicefrac{1}{4}$ \\
    $\ei(y)$    & 0.42 & - & 2.00 & -          & 0.42 & - & - & 2.00 \\
    $\phi(y)$   & 0.33 & - & 1.00 & -          & 0.25 & - & - & 0.00 \\
\bottomrule
\end{tabular}
	\caption{Examples AND-ZERO and AND-AND show that $\phi(y)$ sometimes \emph{decreases} with duplicate computation.  Here, the highest $\phi$ of AND-AND is \emph{less} than the lowest $\phi$ of AND-ZERO.  This carries into the average case with AND-ZERO's $\bphi=0.5$ and AND-AND's $\bphi=0.189$ bits.  A dash means that particular $y$ is unreachable for the network.}
	\label{fig:duplicates}
\end{figure}

\textbf{$\phi$ sometimes decreases with duplicate computation.}  In \figref{fig:duplicates} we take a simple system, AND-ZERO, and duplicate the AND node yielding AND-AND.  We see the two systems remain exceedingly similar.  Both have $\ent{X}=2$ and $\info{X}{Y}=0.811$ bits.  Likewise, both have two $Y$ states occurring with probability $\nicefrac{3}{4}$ and $\nicefrac{1}{4}$ giving $\ei(X \to y)$ equal to $0.42$ and $2.00$ bits respectively.  However, their $\phi$ values are quite different.

Only knowing that the $\phi$'s for AND-AND and AND-ZERO are different, we'd expect AND-AND to be higher because an AND node ``does more'' than a ZERO node (simply shutting off). But instead we get the opposite---AND-AND's highest $\phi$ is \emph{less} than AND-ZERO's lowest $\phi$!  The ideal measure of integrated information might be invariant or increase under duplicate computation, but it certainly wouldn't decrease.

\begin{figure}[htb]
\centering
	\subfloat[SHIFT]{
	      \includegraphics[width=1.18in]{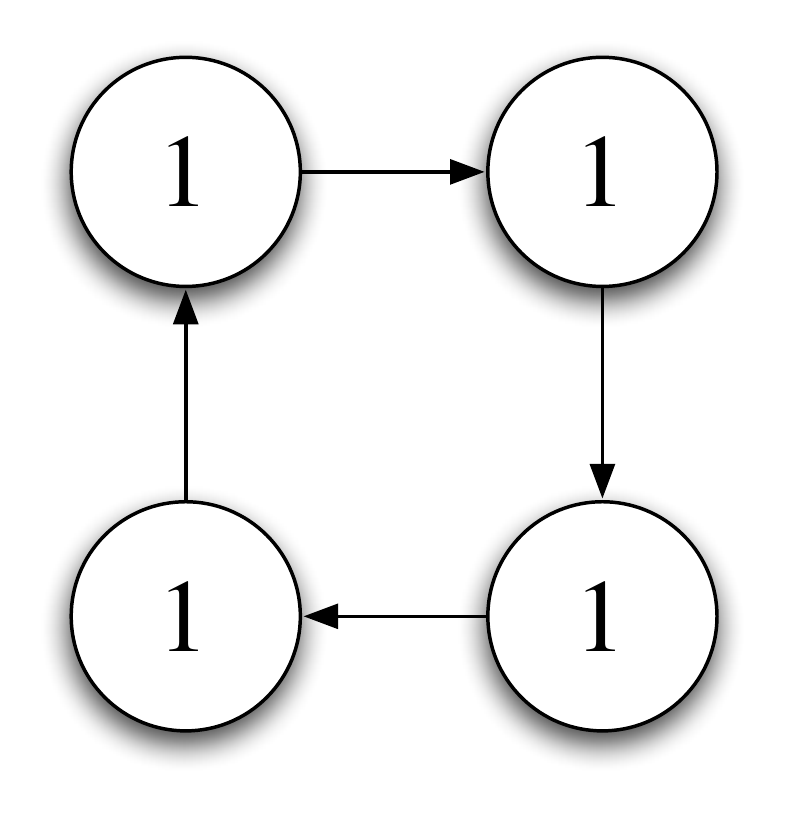}
	       \label{fig:SHIFT}
	       }
	\subfloat[4422]{
	      \includegraphics[width=1.3in]{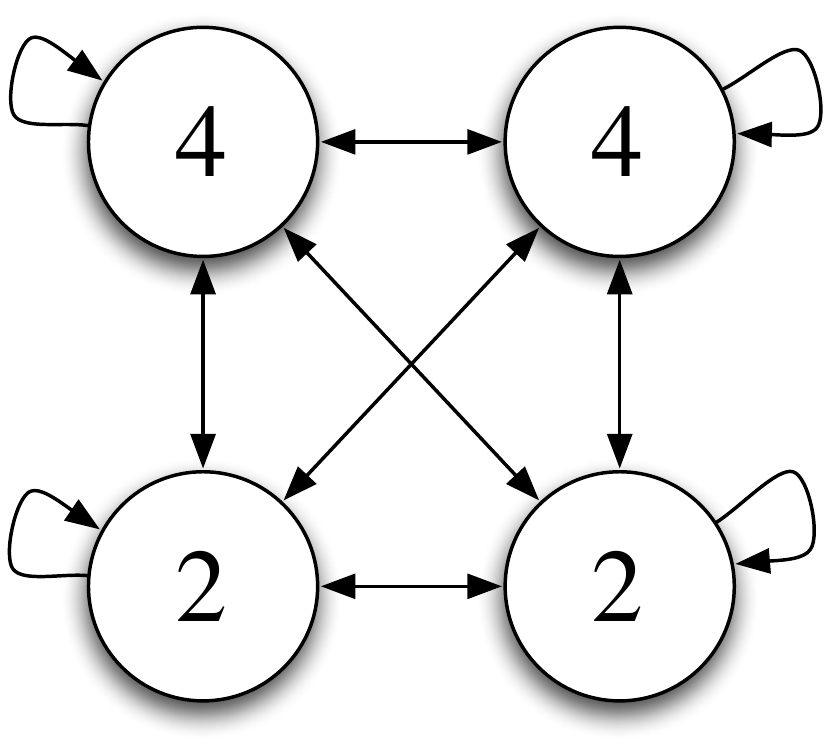}
	       \label{fig:4422}
	       }
	\subfloat[4322]{
	      \includegraphics[width=1.3in]{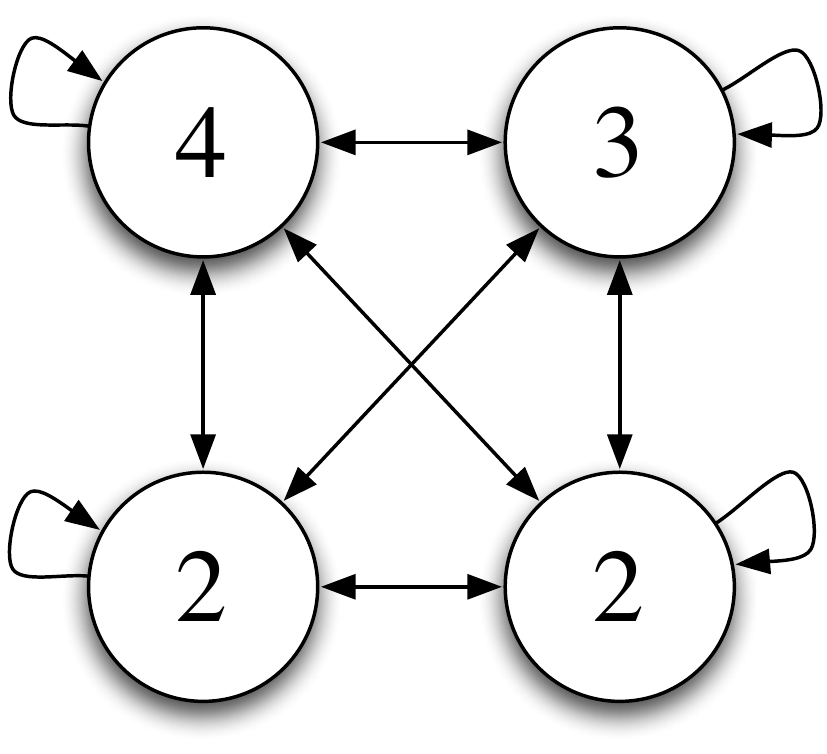}
	       \label{fig:4322}
	       }	       
	\subfloat[4321]{
	      \includegraphics[width=1.3in]{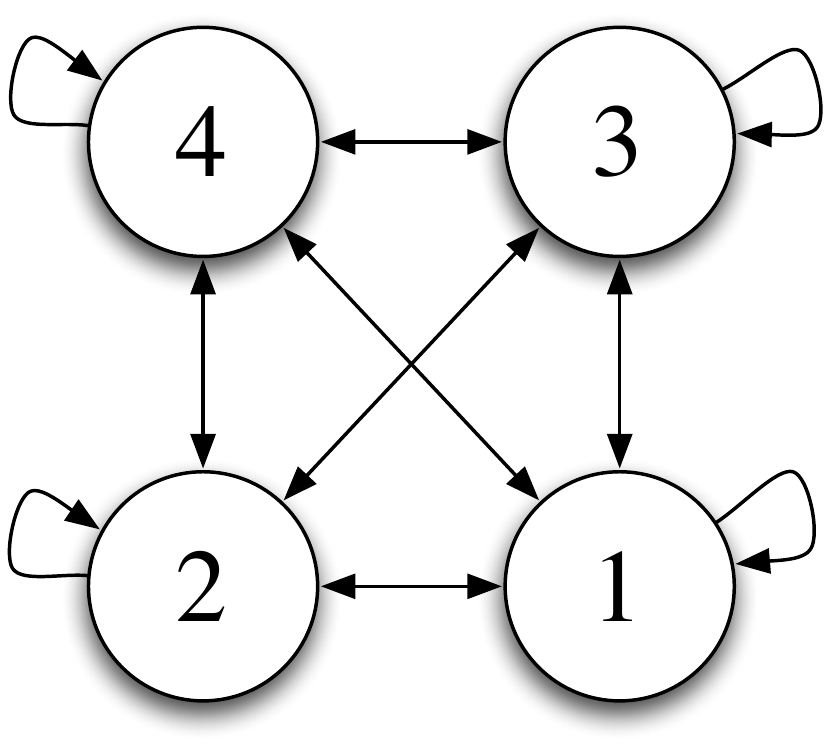}
	       \label{fig:4321}
		   }	
\vspace{0.3in}
\begin{tabular}{l c c c c c c } \toprule
	Network & $\info{X}{Y}$ & $\displaystyle \min_{y} \phi(y)$ & $\displaystyle \max_{y} \phi(y)$ & $\bphi$ \\
\midrule
	SHIFT      & 4.000 & 2.000 & 2.000 & 2.000 \\
	4422       & 1.198 & 0.000 & 0.673 & 0.424 \\
	4322       & 1.805 & 0.322 & 1.586 & 1.367 \\
	4321       & 2.031 & 0.322 & 1.682 & 1.651 \\
\bottomrule
\end{tabular}
\caption{State-dependent $\phi$ and $\bphi$ tell the same story---the $\phi$ value of SHIFT trounces the $\phi$ of the other three networks.  A more intuitive complexity measure would instead increase left from to right.}
\label{fig:diverse}
\end{figure}

\textbf{$\phi$ does not increase with cooperation among diverse parts.}
The $\phi$ measure is sometimes described as corresponding to the juxtaposition of ``functional segregation'' and ``functional integration''.  In a similar vein, $\phi$ is intuited as corresponding to ``interdependence/cooperation among diverse parts''.  \figref{fig:diverse} presents four examples showing that neither intuition is well-captured by the existing $\phi$ measure.

In the first example, SHIFT (\figref{fig:SHIFT}), the state of every node is shifted one-step clockwise---nothing more.  The nodes are homogeneous and each node is wholly determined by its preceding node.  In the three remaining networks (Figures \ref{fig:4422}--\subref*{fig:4321}), every node is a function of all nodes in the network (including itself).  This is to maximize the interdependence/cooperation among the nodes for high ``functional integratation''.  Having established high cooperation, we increase the diversity or ``functional segregation'' from \figref{fig:4422} to \ref{fig:4321}.

By the former intuitions, we'd expect SHIFT (\figref{fig:SHIFT}) to have the lowest $\phi$ and 4321 (\figref{fig:4321}) to have the highest.  But this is not the case.  Instead, SHIFT, the network with the \emph{least} cooperation (every node is a function of one other) and the \emph{least} diverse mechanisms (all nodes have threshold 1) has a $\phi$ far exceeding the others---SHIFT's lowest $\phi$ value at two bits dwarfs even the highest $\phi$ values in Figures \ref{fig:4422}--\subref*{fig:4321}.

SHIFT having the highest integrated information is unexpected, but it's not outright absurd.  SHIFT does have the highest mutual information $\info{X}{Y}$---so the information part is solid.  Is SHIFT integrated?  Well, in SHIFT each node is wholly determined by an external force (the preceding node); so SHIFT is ``integrated'' for a sense of the term.  Whether it makes sense for SHIFT to have the highest integrated information ultimately comes down to precisely what is meant by the term ``integration''.  But even accepting that SHIFT is in some sense integrated, example 4321 is integrated for a palpably stronger sense of the term.  Therefore, until there's an argument that the form of integration present in SHIFT is sufficient for awareness, from a purely theoretical perspective it makes sense to prefer 4321 over SHIFT.

\section{A Novel Measure of Irreducibility to a Partition}
\label{section:statedeppsi}
Our proposed measure $\psi$ quantifies the magnitude of information in $\info{X}{y}$ (eq.~\eqref{eq:EIspecificsurprise}) that is irreducible to a partition of the system at time=$0$.  We define our measure as,
\begin{equation}
    \psi( \setX : y ) \equiv \info{X}{y} - \max_{\mathbf{P}}\Icupe{X^\mathbf{P}_1, \ldots, X^\mathbf{P}_m}{y} \; ,
\label{eq:psidef}
\end{equation}

where $\mathbf{P}$ enumerates over all partitions of set \setX, and $\Icup$ is the information about state $y$ conveyed by the ``union'' across the $m$ parts at time=$0$.  To compute the union information $\Icup$ we use the Partial Information Decomposition (PID) framework.  In PID, $\Icup$ is the inclusion–exclusion dual of $\Icap$.  Thus we can express $\Icup$ solely in terms of $\Icap$ by,
\[
\Icupe{X^\mathbf{P}_1, \ldots, X^\mathbf{P}_m}{y} = \sum_{\mathbf{S} \subseteq \left\{X^\P_1, \ldots, X^\P_m\right\} } (-1)^{\left|\mathbf{S}\right|+1} \Icape{S_1, \ldots, S_{|\mathbf{S}|}}{y} \; .
\]

Conceptually, the intersection information $\Icape{S_1, \ldots, S_{|\mathbf{S}|}}{y}$ quantifies the magnitude of the ``same information'' about state $y$ conveyed by each $S_1, \ldots, S_{|\mathbf{S}|}$.  Although there's currently some debate\cite{bertschinger12,Iwedge} about what is the best $\Icap$ measure, there's consensus that the intersection information $n$ arbitrary random variables $Z_1, \ldots, Z_n$ carry about state $y$ must satisfy the following properties:

\begin{itemize}
    \item[\GP] Global Positivity: $\Icape{Z_1,\ldots,Z_n}{y} \geq 0$.

    \item[\Szero] Weak Symmetry:
$\Icape{Z_1,\ldots,Z_n}{y}$ is invariant under reordering $Z_1, \ldots, Z_n$.

    \item[\SR] Self-Redundancy: $\Icape{Z_1}{y} = \info{Z_1}{y} = \DKL{\Prob{Z_1|y}}{\Prob{Z_1}}$. The intersection information a single predictor $Z_1$ conveys about the target state $y$ is equal to the ``specific surprise''\cite{meister-99}.

    \item[\Mone] Strong Monotonicity:  $\Icape{Z_1,\ldots,Z_n, W}{y} \leq \Icape{Z_1,\ldots,Z_n}{y}$ with equality if there exists $Z_i \in \{Z_1, \ldots, Z_n\}$ such that $\info{W Z_i}{y} = \info{W}{y}$ where $W Z_i$ is the joint random variable (cartesian product) of $W$ and $Z_i$.

    \item[\Eq] Equivalence-Class Invariance:
$\Icape{Z_1,\ldots,Z_n}{y}$ is invariant under substituting $Z_i$ (for any $i=1,\ldots,n$) by an informationally equivalent random variable\cite{Iwedge}.\footnote{Meaning $\Icap$ is invariant under substituting $Z_i$ with $W$ if $\ent{Z_i \middle| W} = \ent{W \middle| Z_i} = 0$.}  Similarly, $\Icape{Z_1, \ldots, Z_n}{y}$ is invariant under substituting state $y$ for state $w$ if $\Prob{w|y}=\Prob{y|w} = 1$.

\end{itemize}

Now we take a less common course---instead of choosing a particular $\Icap$ that satisfies the above properties, we will simply use the properties above directly to bound the range of possible $\psi$ values.  Leveraging \Mone, \Szero, and \SR, eq.~\eqref{eq:psidef} simplifies to,\footnote{See Appendix \ref{proof:bipartitions} for a proof.}
\begin{equation}
\begin{split}
    \psi( \setX : y) &= \info{X}{y} - \max_{A \subset \setX} \Icupe{A,B}{y} \\
    &= \info{X}{y} - \max_{A \subset \setX} \left[ \info{A}{y} + \info{B}{y} - \Icape{ A, B }{y} \right] \; ,
\end{split}
\label{eq:psisimplified}
\end{equation}
where $A \not= \emptyset$ and $B \equiv \setX \setminus A$.

From eq.~\eqref{eq:psisimplified}, the only undefined term is $\Icape{A,B}{y}$.  Leveraging \GP, \Mone, and \SR, we can bound it by,
\begin{equation}
\label{eq:dispIcapbounds}
    0 \leq \Icape{A, B}{y} \leq \min \left[ \info{A}{y}, \info{B}{y} \right] \; .
\end{equation}

Finally, we bound $\psi$ by plugging in the above bounds on $\Icape{A,B}{y}$ into eq.~\eqref{eq:psisimplified}.  With some algebra and leveraging assumption \textbf{(B)}, this yields the following bounds for $\psi$,\footnote{See Appendix \ref{appendix:psibounds} for proofs.} 
\begin{equation}
\begin{split}
    \psi_{\min}\!\left(\setX : y \right) &= \min_{A \subset \setX} \DKL{ \Prob{X\middle|y} }{ \Prob{A|y} \Prob{B|y} } \\
    \psi_{\max}\!\left(\setX :y \right) &= \min_{i \in \{1, \ldots, n\}} \DKL{ \Prob{X \middle|y} }{ \Prob{X_i} \Prob{X_{\sim i} \middle| y} } \; ,
\end{split}
\label{eq:psiminmax}
\end{equation}
where $X_{\sim i}$ is the random variable of all nodes in $X$ excluding node $i$.  Then, \newline $\psi_{\min}( \setX : y ) \leq \psi( \setX : y ) \leq \psi_{\max}( \setX : y )$.

\subsection{Stateless $\psi$ is $\bpsi$}

We define $\bpsi$ analogous to $\phi$ per Section \ref{section:bracketphi}.  To compute $\bpsi$ we weaken the properties in Section \ref{section:statedeppsi} so that they only apply to the average case, i.e., the properties \GP, \Mone, \Szero, \SR, and \Eq don't have to apply for each $\Icape{Z_1, \ldots, Z_n}{y}$, but merely for the average case $\Icape{Z_1, \ldots, Z_n}{Y}$.

Via the same algebra from eq.~\eqref{eq:psisimplified}, $\bpsi$ simplifies to,
\begin{equation}
\begin{split}
    \bpsi(X_1, \ldots, X_n : Y) &\equiv \info{X}{Y} - \max_{\mathbf{P}}\Icupe{X^\P_1, \ldots, X^\P_m}{Y} \\
    &= \info{X}{Y} - \max_{A \subset \setX} \Icupe{A,B}{Y} \\
    &= \info{X}{Y} - \max_{A \subset \setX} \left[ \info{A}{Y} + \info{B}{Y} - \Icape{ A, B }{Y} \right]\; ,
\end{split}
\end{equation}
where $A \not= \emptyset$ and $B \equiv \setX \setminus A$.  Using the weakened properties, we likewise have $0 \leq \Icape{A,B}{Y} \leq \min\left[ \info{A}{Y}, \info{B}{Y} \right]$.  Plugging in these $\Icap$ bounds yields the following bounds $\bpsi$,\footnote{See Appendix \ref{appendix:bpsibounds} for proofs.} 
\begin{equation}
\begin{split}
    \bpsi_{\min}\!\left( \setX : Y \right) &= \min_{A \subset \setX} \info{A}{B\middle|Y} \\
    \bpsi_{\max}\!\left( \setX : Y \right) &= \min_{i \in \{1, \ldots, n\}} \DKL{ \Prob{X, Y} }{ \Prob{X_{\sim i}, Y} \Prob{X_i} } \; ,
\end{split}
\label{eq:bpsiminmax}
\end{equation}
where $X_{\sim i}$ is the random variable of all nodes in $X$ excluding node $i$.  Then, \newline $\bpsi_{\min}( \setX : Y ) \leq \bpsi( \setX : Y ) \leq \bpsi_{\max}( \setX : Y)$.

\section{Contrasting $\psi$ versus $\phi$}

\textbf{Theoretical benefits of $\psi$.} The overarching theoretical benefit is that $\psi$ is entrenched within the rigorous Partial Information Decomposition framework\cite{plw-10}.  PID builds a principled irreducibility measure from a redundancy measure $\Icap$.  Here we only take the most accepted properties of $\Icap$ to bound $\psi$ from above and below.  As the complexity community converges on the additional properties $\Icap$ must satisfy\cite{bertschinger12,Iwedge}, the derived bounds on $\psi$ will tighten.

There are four benefits of $\psi$'s principled underpinning.  First, whereas $\phi(y)$ can exceed the entropy of the whole system, i.e., $\phi(y) \not \leq \ent{X}$, $\psi(y)$ is bounded by specific-surprise, i.e., $\psi(y)~\leq~\info{X}{y} = \DKL{ \Prob{X|y} }{ \Prob{X} }$.  This gives $\psi$ the natural info-theoretic interpretation for the state-dependent case which $\phi$ lacks.  Second, PID provides justification for $\psi$ not needing a MIP normalization and thus eliminates a longstanding ambiguity about $\phi$\cite{balduzzi_private}.  Third, PID is a flexible framework that enables quantifying irreducibility to overlapping parts should we decide to explore it.\footnote{Unlike disjoint parts, the maximum union information over two overlapping parts is not equal to the maximum union information over $m$ overlapping parts.  See \cite{qirred} for two measures of irreducibility to overlapping parts.}

One final perk is that $\psi$ is already substantially faster to compute.  Whereas computing $\phi$ scales\footnote{This comes from eq.~\eqref{eq:phiMIP} enumerating all partitions (Bell's number) of $n$ elements.} $O(n!)$, computing $\psi$ scales\footnote{This comes from eq.~\eqref{eq:psisimplified} enumerating all $2^{n-1} - 1$ bipartitions of $n$ elements.} $O(2^n)$---a substantial improvement that may improve even further as the complexity community converges on additional properties of $\Icap$.

\begin{figure}[h!]
\centering

\subfloat[AND-ZERO+KEEP]{\includegraphics[width=1.5in]{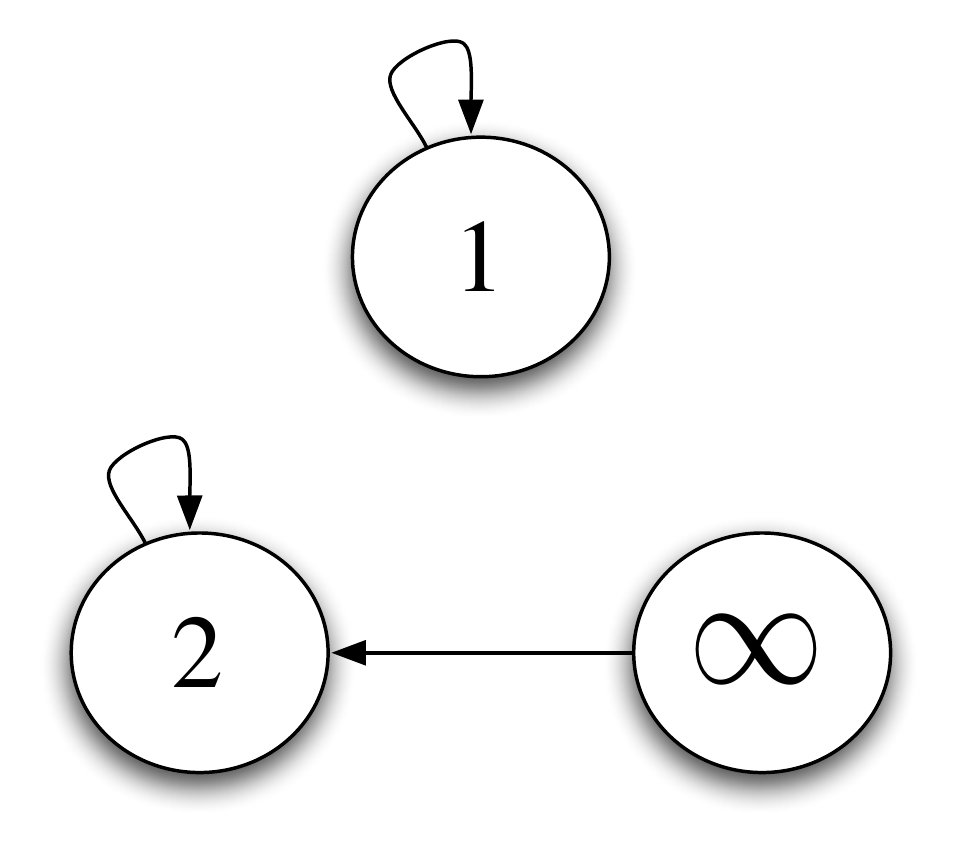} \label{fig:AZK} }
\subfloat[2x AND-ZERO]{\includegraphics[width=1.5in]{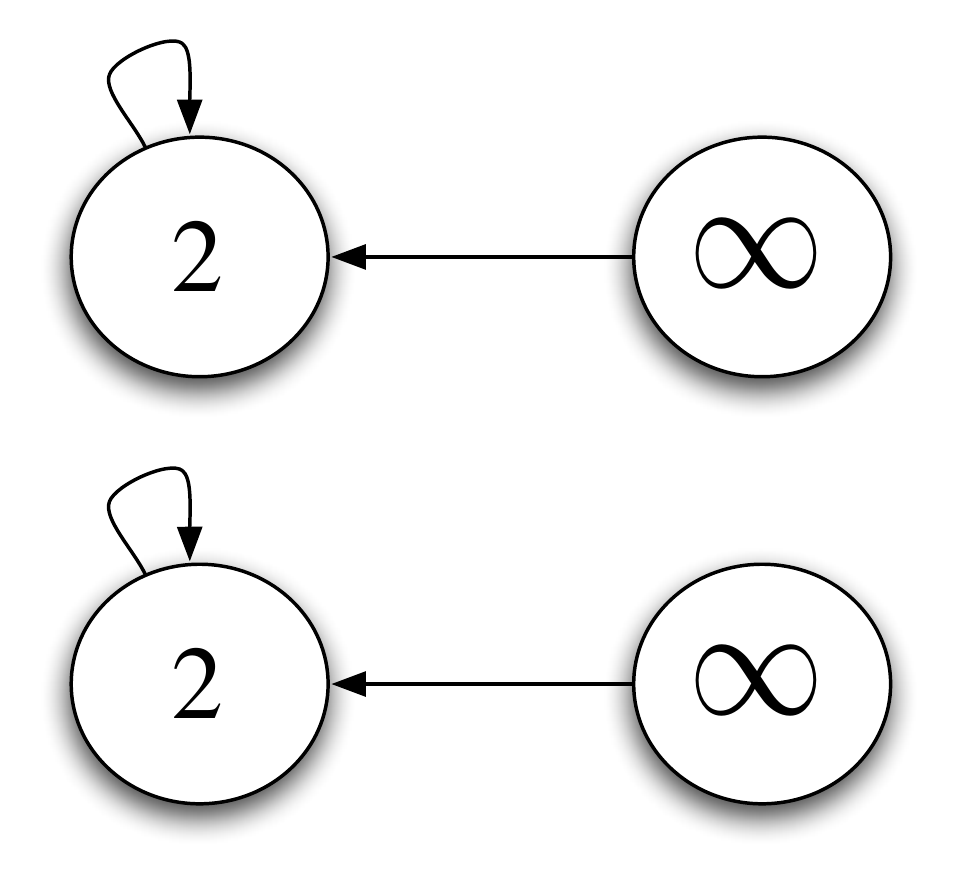} \label{fig:AZ_AZ} }	
	\subfloat[KEEP-KEEP]{ \includegraphics[width=1.5in]{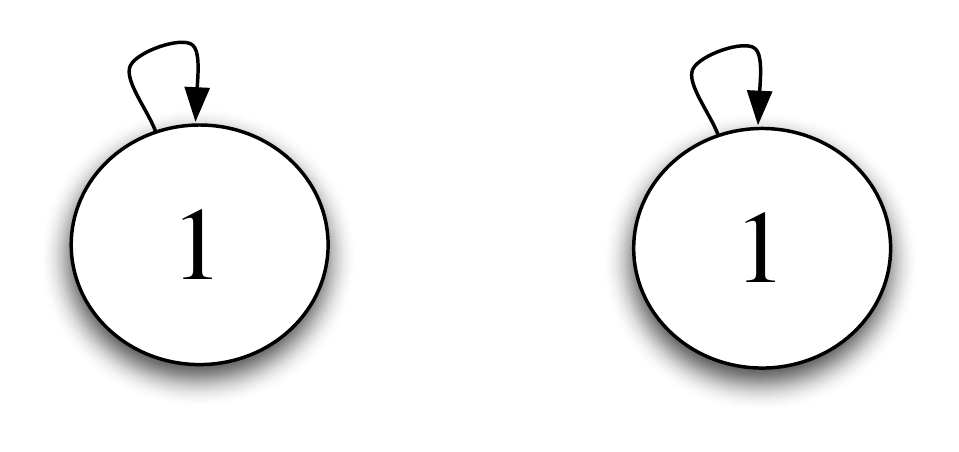} \label{fig:KK_one}  }
	
	\subfloat[GET-GET]{ \includegraphics[width=1.5in]{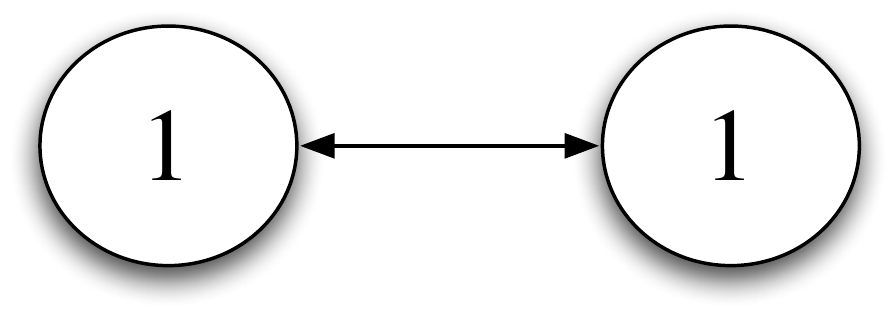} \label{fig:GG_one} }
	\subfloat[ANDtriplet]{ \includegraphics[width=1.5in]{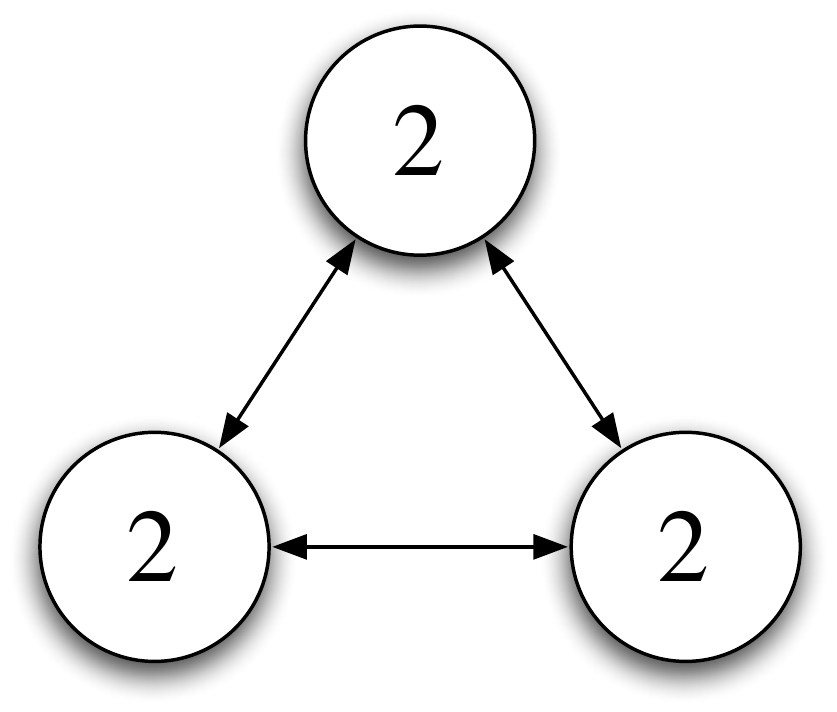} \label{fig:AAA_one} }
	\subfloat[iso-ANDtriplet]{ \includegraphics[width=1.5in]{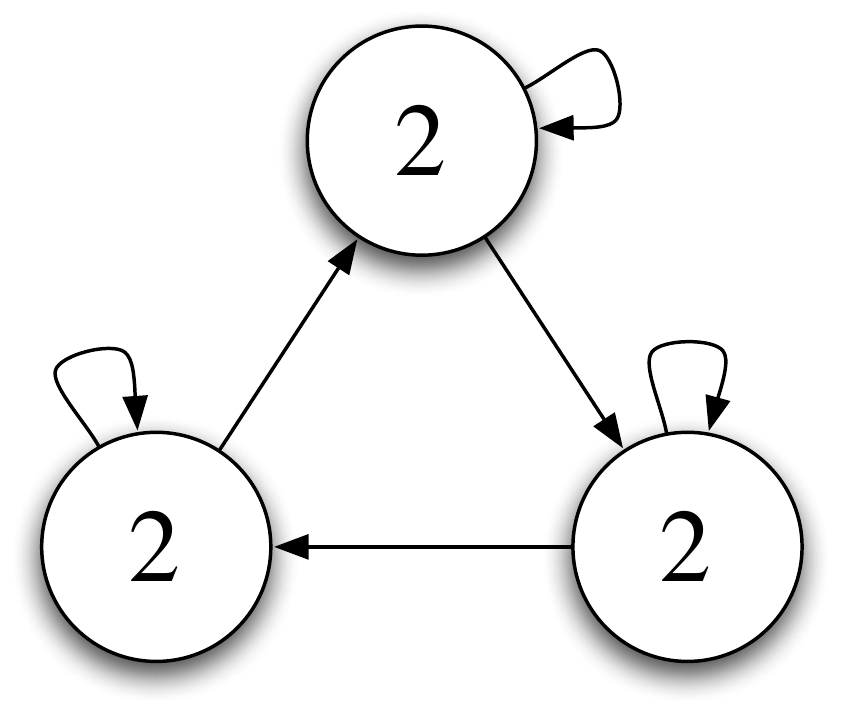} \label{fig:rArArA_one} }
	
    \subfloat[AND-ZERO]{ \includegraphics[width=1.5in]{AZ.pdf} \label{fig:AZ_one} }
	\subfloat[AND-AND]{ \includegraphics[width=1.5in]{AA.pdf} \label{fig:AA_one}  }
	
\vspace{0.2in}

\begin{tabular}{  l l l l l l } \toprule
 \addlinespace
		Network & $\info{X}{Y}$ & $\bphi$ & $\bpsi_{\min}$ & $\bpsi_{\max}$ \\
	\midrule
	AND-ZERO+KEEP \subref{fig:AZK} & 1.81 & 0 & 0 & 0.50 \\
	2x AND-ZERO \subref{fig:AZ_AZ} & 1.62 & 0 & 0 & 0.50 \\
	\addlinespace
	KEEP-KEEP \subref{fig:KK_one} & 2.00 & 0 & 0 & 1.00 \\
	GET-GET \subref{fig:GG_one} & 2.00 & 2.00 & 0 & 1.00 \\    
    \addlinespace
	ANDtriplet \subref{fig:AAA_one} & 2.00 & 2.00 & 0.16 & 0.75 \\
	iso-ANDtriplet \subref{fig:rArArA_one} & 2.00 & 1.07 & 0.16 & 0.75  \\
    \addlinespace    
	AND-ZERO \subref{fig:AZ_one}  & 0.81 & 0.50 & 0.19 & 0.50 \\
	AND-AND \subref{fig:AA_one} & 0.81 & 0.19 & 0.19 & 0.50 \\
	\addlinespace	
	SHIFT (Fig. \ref{fig:SHIFT}) & 4.00 & 2.00 & 0 & 1.00 \\
	4422 (Fig. \ref{fig:4422})   & 1.20 & 0.42 & 0.33 & 0.50 \\
	4322 (Fig. \ref{fig:4322})   & 1.81 & 1.37 & 0.68 & 0.88 \\
	4321 (Fig. \ref{fig:4321})   & 2.03 & 1.65 & 0.78 & 1.00 \\
	\bottomrule	
	\end{tabular}

\caption{Contrasting $\bphi$ versus $\bpsi$ for exemplary networks.}
\label{fig:necessary}
\end{figure}

\textbf{Behavioral differences between $\psi$ and $\phi$.}
The first row in \figref{fig:necessary} shows two ways a network can be irreducible to atomic elements (the nodes) yet still reducible to disjoint parts.  Compare AND-ZERO (\figref{fig:AZ_one}) to AND-ZERO+KEEP (\figref{fig:AZK}).  Although AND-ZERO is irreducible, AND-ZERO+KEEP reduces to the bipartition separating the AND-ZERO component and the KEEP node.  This reveals how fragile measures like $\psi$ and $\phi$ are---add a single disconnected node and they plummet to zero.  Example 2x AND-ZERO (\figref{fig:AZ_AZ}) shows that a wholly reducible network can be composed entirely of irreducible parts.

Example KEEP-KEEP (\figref{fig:KK_one}) highlights the only known relative drawback of $\psi$---$\psi$'s current upperbound is painfully loose.\footnote{The current upperbounds are $\psi_{\max}$ in eq.~\eqref{eq:psiminmax} and $\bpsi_{\max}$ in eq.~\eqref{eq:bpsiminmax}.}  The desired irreducibility for KEEP-KEEP is zero bits, and indeed, $\psi_{\min}$ is $0$ bits---but $\psi_{\max}$ is a monstrous $1$ bit!  We rightly expect tighter bounds for such easy examples like KEEP-KEEP.  Tighter bounds on $\Icap$ (and thus $\psi$) is an area of active research but as-is the bounds are loose.

Example GET-GET (\figref{fig:GG_one}) epitomizes the most striking difference between $\psi$ and $\phi$.  By property \Eq, the $\psi$ values for KEEP-KEEP and GET-GET are provably equal (making the desired $\psi$ for GET-GET zero bits), yet their $\phi$ values couldn't be more different.  Although the $\phi$ for KEEP-KEEP is zero, the $\phi$ for GET-GET is the maximal (!) two bits of irreducibility.  Whereas $\psi$ views GET nodes as non-integrative, $\phi$ views GET nodes as maximally integrative.

This begs the question---should GETs be integrative?  It's sensible for GETs to be mildly integrative, but the logic of partitioning the system forces us to choose between GETs being non-integrative (akin to a KEEP) or maximally integrative.  To resolve this dilemma this we return to \figref{fig:diverse}.  The primary benefit of $\psi$ making KEEPs and GETs equivalent is that $\psi$ is zero for chains of GETs such as the SHIFT network (\figref{fig:SHIFT}).  This enables $\psi$ to better match our intuition for ``cooperation among diverse parts''.  For example, in \figref{fig:diverse} the network with the highest $\phi$ is the counter-intuitive SHIFT, but the network with the highest $\psi$ is the more sensible 4321 (see table in \figref{fig:necessary}).  With these examples in mind, we personally believe GETs being non-integrative is the better choice.

The third row in \figref{fig:necessary} shows how $\psi$ and $\phi$ respectively treat self-connections.  In ANDtriplet (\figref{fig:AAA_one})  and iso-ANDtriplet (\figref{fig:rArArA_one}) each node integrates information about two nodes.  The only difference is that in ANDtriplet each node integrates information about two \emph{other} nodes, while in iso-ANDtriplet each node integrates information is about \emph{itself} and one other.

Just as $\psi$ views KEEP and GET nodes equivalently, $\psi$ views self and cross connections equivalently.  In fact, by property \Eq the $\psi$ values for ANDtriplet and iso-ANDtriplet are provably equal.    Alternatively, $\phi$ considers self and cross connections differently in that $\phi$ can only decrease when adding a self-connection.  As such, the $\phi$ for iso-ANDtriplet is less than ANDtriplet.

The fourth row in \figref{fig:necessary} shows this same self-connections business carrying over to duplicate computations.  Although AND-AND (\figref{fig:AA_one}) and AND-ZERO (\figref{fig:AZ_one}) perform the same computation, AND-AND has an additional self-connection that pushes AND-AND's $\phi$ below that of AND-ZERO.  By \Eq, the $\psi$ values of AND-ZERO and AND-AND are provably equal.

\section{Conclusion}
Regardless of any connection to consciousness, purely as a measure of functional irreducibility we have three concerns about $\phi$: (1) state-dependent $\phi$ can exceed the entropy of the entire system; (2) $\phi$ often decreases with duplicate computation; (3) $\phi$ doesn't match the intuition of ``cooperation among diverse parts''.

We introduced a new irreducibility measure, $\psi$, that solves all three concerns but otherwise stays close to the original spirit of $\phi$---i.e., the quantification of a system's irreducibility to disjoint parts.  Based in Partial Information Decomposition, $\psi$ has other desirable properties such as not needing a MIP normalization and being substantially faster to compute.  We then contrasted $\psi$ versus $\phi$ in binary networks.

Although we endorse $\psi$ over $\phi$, the $\psi$ measure remains imperfect.  The most notable areas for improvement are:
\begin{enumerate}
    \item The current $\psi$ bounds are too loose.  We need to tighten the $\Icap$ bounds (eq.~\eqref{eq:dispIcapbounds}), which will tighten the derived bounds on $\psi$ and $\bpsi$.
    \item Justify why a measure of conscious experience should prefer irreducibility to disjoint parts over irreducibility to overlapping parts.
    \item Reformalize the work on qualia in \cite{balduzzi09} using $\psi$ or comparable measure.
    \item Although not specific to $\psi$, there needs to be a stronger justification for the chosen distribution on $X$ (see Appendix \ref{appendix:Xdistribution}).
\end{enumerate}

Our introduced $\psi$ measure effortlessly generalizes to the quantum case simply by replacing all instances of Shannon mutual information in eq.~\eqref{eq:psisimplified} with von Neumann (quantum) information.  This ``quantum $\psi$'' is a quantum infotheoretic measure that remains much more faithful to its parents \cite{balduzzi-tononi-08,tononi-08} than Tegmark's innovative perceptronium implementation\cite{tegmark14}.

\bibliography{quant_synergy}

\appendix
\part*{Appendix}
\section{Reading the Network Diagrams}
\label{app:tutorial}
\begin{figure}[h!]
\centering
	\subfloat[ZERO-ZERO]{ \includegraphics[width=1.33in]{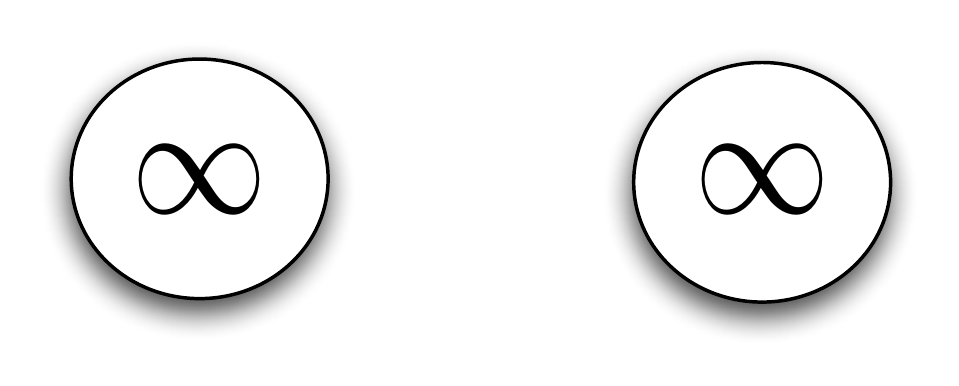} \label{fig:ZZ} }
	\subfloat[KEEP-ZERO]{ \includegraphics[width=1.33in]{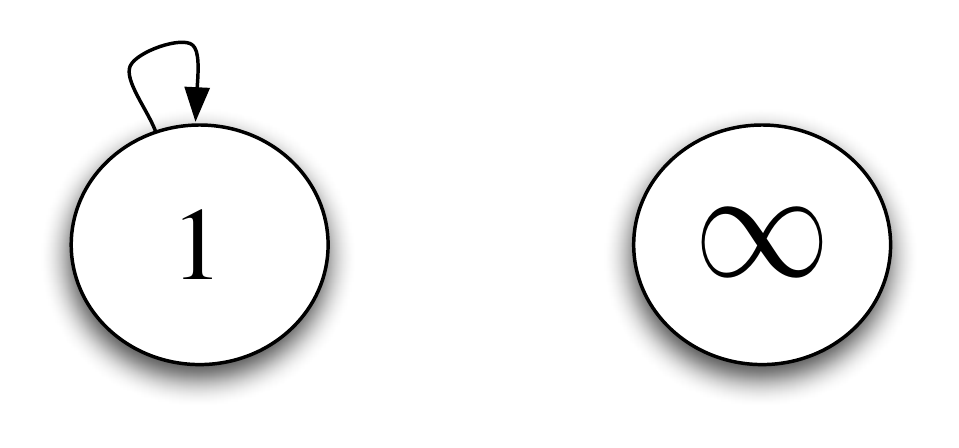} \label{fig:KZ} }
	\subfloat[GET-ZERO]{ \includegraphics[width=1.33in]{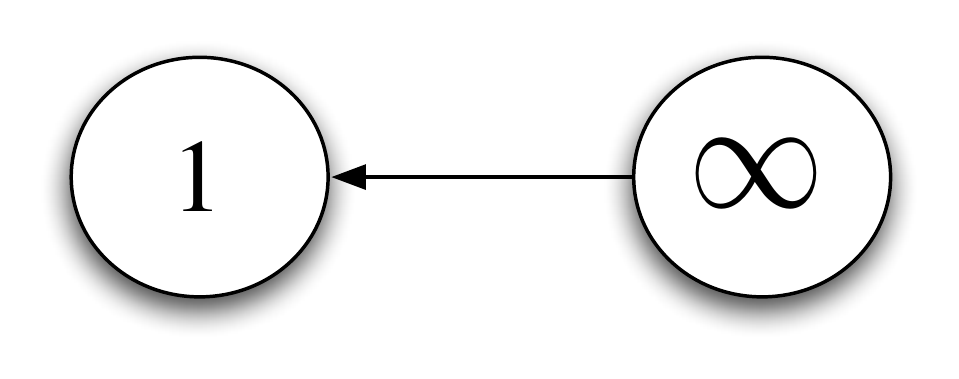} \label{fig:GZ} }
	\subfloat[KEEP-KEEP]{ \includegraphics[width=1.33in]{KK.pdf} \label{fig:KK} }	
	
	\subfloat[GET-KEEP]{ \includegraphics[width=1.33in]{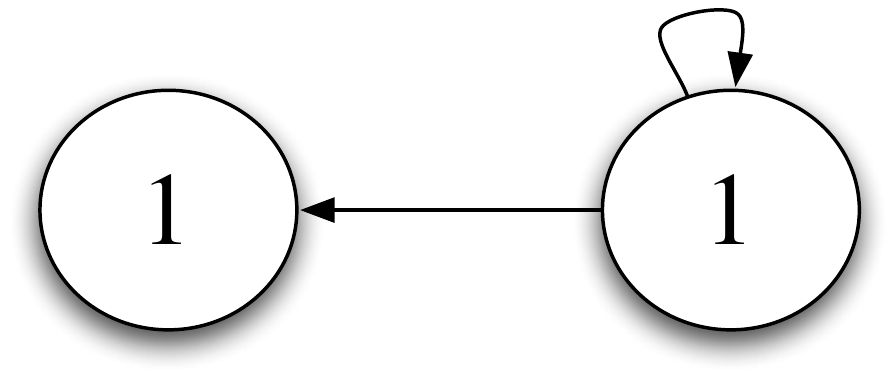} \label{fig:GK} }	
	\subfloat[GET-GET]{ \includegraphics[width=1.33in]{GG.pdf} \label{fig:GG} }	
	\subfloat[AND-ZERO]{ \includegraphics[width=1.33in]{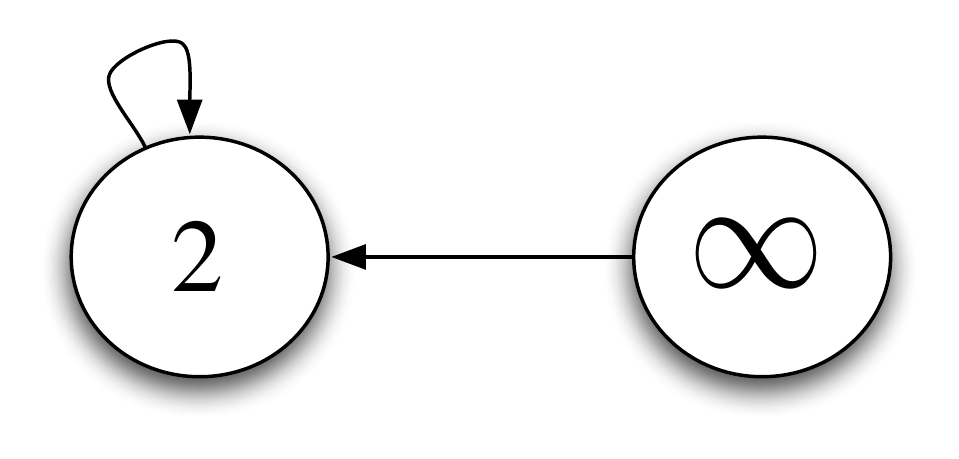} \label{fig:AZ_app} }		
	\subfloat[AND-XOR]{ \includegraphics[width=1.33in]{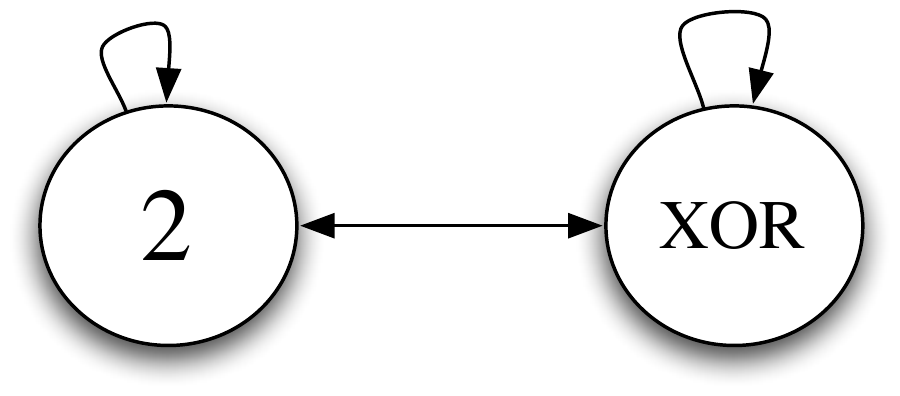} \label{fig:AX} }			

\vspace{0.2in}
\begin{tabular}{c c c c c c c c c c c} \toprule
	\multirow{2}{*}{$X$} & \ & ZERO- & KEEP- &GET- & KEEP- &GET-&GET-&AND-&AND-\\
	 & \ & ZERO& ZERO & ZERO & KEEP & KEEP & GET &ZERO & XOR \\
\midrule
	\bin{00} & $\to$ & \bin{00} & \bin{00} & \bin{00} & \bin{00} & \bin{00} & \bin{00} & \bin{00} & \bin{00} \\
	\bin{01} & $\to$ & \bin{00} & \bin{00} & \bin{10} & \bin{01} & \bin{11} & \bin{10} & \bin{00} & \bin{01} \\
	\bin{10} & $\to$ & \bin{00} & \bin{10} & \bin{00} & \bin{10} & \bin{00} & \bin{01} & \bin{00} & \bin{01} \\
	\bin{11} & $\to$ & \bin{00} & \bin{10} & \bin{10} & \bin{11} & \bin{11} & \bin{11} & \bin{10} & \bin{10} \\
	\bottomrule
\end{tabular}

\caption{Eight doublet networks with transition tables.}
\label{fig:tutorial}
\end{figure}

We present eight doublet networks and their transition tables so you can see how the network diagram specifies the transition table.  \figref{fig:tutorial} shows eight network diagrams to build your intuition.  The number inside each node is that node's \emph{activation threshold}.  A node updates to \bin{1} (conceptually an ``ON'') if there at least as many of inputs ON as its activation threshold; e.g. a node with an inscribed 2 updates to a \bin{1} if two or more incoming wires are ON.  An activation threshold of \scalebox{1.15}{$\infty$} means the node always updates to \bin{0} (conceptually an ``OFF'').  A binary string denotes the state of the network, read left to right.

We take the AND-ZERO network (\figref{fig:AZ_app}) as an example.  Although the AND-ZERO network can never output \bin{01} or \bin{11} (Figure 1b), we still consider states \bin{01}, \bin{11} as equally possible states at time=0.  This is because $X$ is uniformly distributed per assumption \textbf{(B)}.

\begin{figure}[h!]
	\subfloat[XOR-ZERO]{ \includegraphics[width=1.2in]{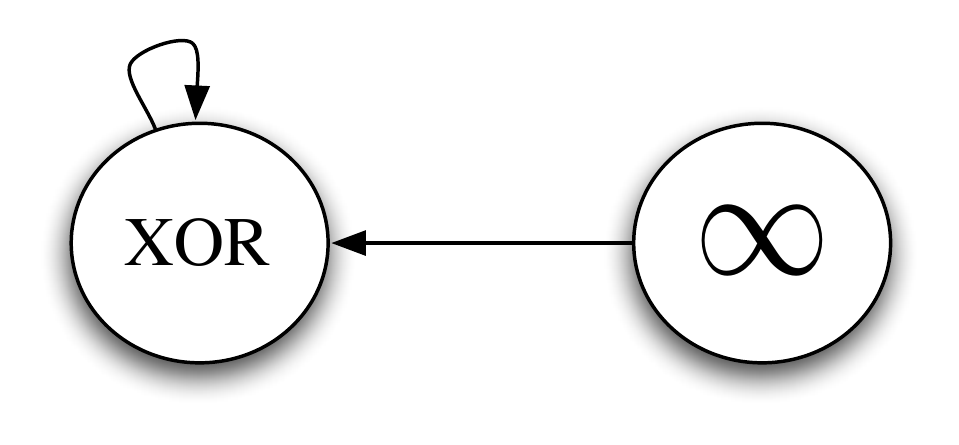} \label{fig:XZ} }
	\subfloat[XOR-KEEP]{ \includegraphics[width=1.2in]{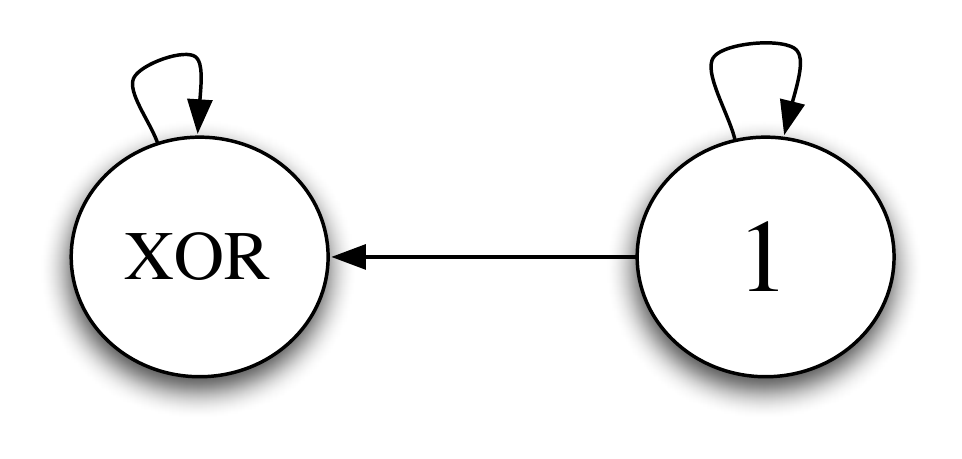} \label{fig:XK} }
	\subfloat[XOR-GET]{ \includegraphics[width=1.2in]{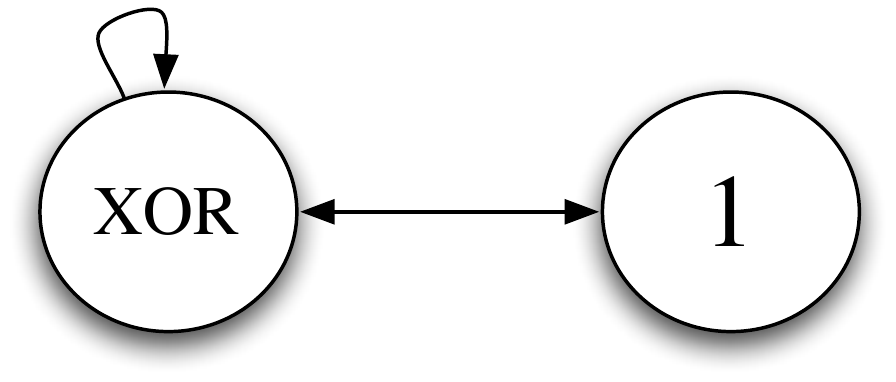} \label{fig:XG} }	
	\subfloat[XOR-XOR]{ \includegraphics[width=1.2in]{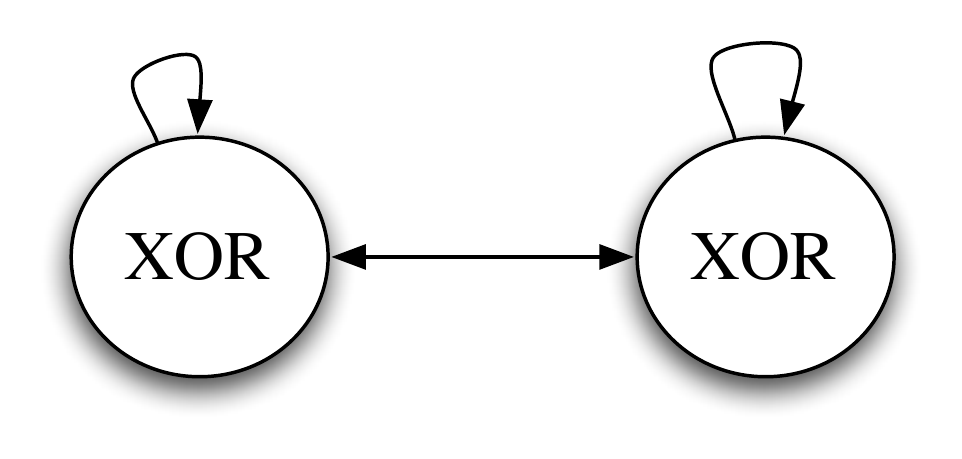} \label{fig:XX} }	

\begin{minipage}{0.3\textwidth}
	\subfloat[XOR-AND]{ \includegraphics[width=1.3in]{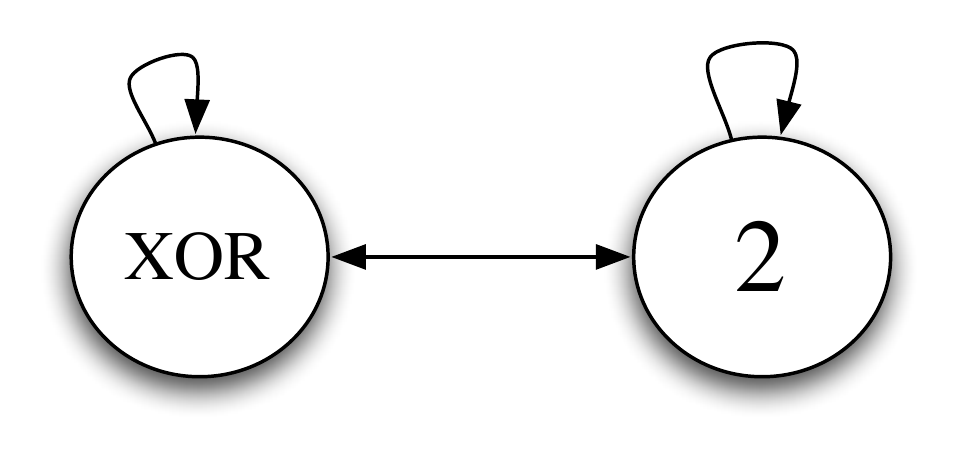} \label{fig:XA} }
\end{minipage}
\begin{minipage}{0.7\textwidth}
\vspace{0.2in}
\begin{tabular}{c c c c c c c c c} \toprule
	\multirow{2}{*}{$X$} & \ & XOR- & XOR- & XOR- & XOR-  & XOR- \\
	& & ZERO & KEEP & GET & XOR & AND \\
\midrule
	\bin{00} & $\to$ & \bin{00} & \bin{00} & \bin{00} & \bin{00} & \bin{00} \\
	\bin{01} & $\to$ & \bin{10} & \bin{11} & \bin{10} & \bin{11} & \bin{10} \\
	\bin{10} & $\to$ & \bin{10} & \bin{10} & \bin{11} & \bin{11} & \bin{10} \\
	\bin{11} & $\to$ & \bin{00} & \bin{01} & \bin{01} & \bin{00} & \bin{01} \\
	\bottomrule
\end{tabular}
\end{minipage}

\centering
\vspace{0.3in}
\begin{tabular}{ l l l l l l } \toprule
 \addlinespace
		Network & $\info{X}{Y}$ & $\bphi$ & $\bpsi_{\min}$ & $\bpsi_{\max}$ \\
	\midrule
	ZERO-ZERO (Fig. \ref{fig:ZZ}) & 0 & 0 & 0 & 0 \\
	KEEP-ZERO (Fig. \ref{fig:KZ}) & 1.0 & 0 & 0 & 0 \\
	KEEP-KEEP (Fig. \ref{fig:KK}) & 2.0 & 0 & 0 & 1.0 \\
	GET-ZERO (Fig. \ref{fig:GZ}) & 1.0 & 1.0 & 0 & 0 \\
	GET-KEEP (Fig. \ref{fig:GK}) & 1.0 & 0 & 0 & 0 \\
	GET-GET (Fig. \ref{fig:GG}) & 2.0 & 2.0 & 0 & 1.0 \\
	\addlinespace
	AND-ZERO (Fig. \ref{fig:AZ}) & 0.811 & 0.5 & 0.189 & 0.5 \\
	AND-KEEP  & 1.5 & 0.189 & 0 & 0.5 \\
	AND-GET   & 1.5 & 1.189 & 0 & 0.5 \\
	AND-AND (Fig. \ref{fig:AA}) & 0.811 & 0.189 & 0.189 & 0.5 \\
	AND-XOR (Fig. \ref{fig:AX}) & 1.5 & 1.189 & 0.5 & 1.0 \\	
	\addlinespace
	XOR-ZERO \subref{fig:XZ} & 1.0 & 1.0 & 1.0 & 1.0 \\
	XOR-KEEP  \subref{fig:XK} & 2.0 & 1.0 & 0 & 1.0 \\
	XOR-GET \subref{fig:XG} & 2.0 & 2.0 & 0 & 1.0 \\
	XOR-AND \subref{fig:XA} & 1.5 & 1.189 & 0.5 & 1.0 \\
	XOR-XOR  \subref{fig:XX} & 1.0 & 1.0 & 1.0 & 1.0 \\
	
	\bottomrule	
	\end{tabular}
\caption{Networks, transition tables, and measures for the diagnostic doublets.}
\label{fig:doublets}
\end{figure}

\clearpage

\section{Necessary Proofs}

\subsection{Proof that Max Union of Bipartitions Covers All Partitions}
\label{proof:bipartitions}
{\lem{Given properties \Szero and \Mone, the maximum union information conveyed by a partition of predictors $\setX = \{X_1, \ldots, X_n\}$ about state $y$ equals the maximum union information conveyed by a bipartition of $\setX$ about state $y$.}}

\begin{proof}
We prove that the maximum information conveyed by a Partition, $\IbP( \setX : y )$, equals the maximum information conveyed by a Bipartition, $\IbPTwo( \setX : y )$ by showing,
\begin{equation}
    \IbP( \setX : y ) \leq \IbPTwo( \setX : y ) \leq \IbP( \setX : y ) \; .
\label{eq:IbBtobeproven}
\end{equation}

First we show that $\IbPTwo( \setX : y) \leq \IbP( \setX : y)$.  By their definitions.
\begin{align*}
    \IbP( \setX : y ) &\equiv \max_{\setP} \Icupe{\setP}{y} \\
    \IbPTwo( \setX : y ) &\equiv \max_{ \substack{\setP \\ |\setP| = 2} } \Icupe{\setP}{y} \; ,    
\end{align*}
where $\setP$ enumerates over all partitions of set $\setX$.

By removing the restriction that $|\setP|=2$ from the maximization in $\IbPTwo$ we arrive at $\IbP$.    As removing a restriction can only increase the maximum, thus $\IbPTwo( \setX : y)~\leq~\IbP(\setX : y)$.

\medskip

Next we show that $\IbP( \setX : y ) \leq \IbPTwo\left( \setX : y \right)$.  Meaning we must show that, 

\begin{equation}
    \max_{\setP}  \; \Icupe{\setP}{y} \leq \max_{ \substack{\setP \\ |\setP| = 2} } \Icupe{\setP}{y} \; .
\label{eq:IbPunion1}    
\end{equation}

Without loss of generality, we choose an arbitrary subset/part $S \subset \setX$.  This yields the bipartition of parts $\{S, \setX \setminus S\}$.  We then further partition the second part, $\setX \setminus S$, into $k$ (disjoint) subparts denoted $T_1, \ldots, T_k$ where $2 \leq k \leq n - \left| S \right|$ creating an arbitrary partition $\mathbf{P}=\{S, T_1, \ldots, T_k\}$.  We now need to show that,
\[
    \Icupe{S, T_1, \ldots, T_k}{y} \leq \Icupe{S, \setX \setminus S}{y} \; .
\]

By \Mone equality condition, we can append each subcomponent $T_1, \ldots, T_k$ to $\{S, \setX \setminus S\}$ without changing the union-information because for each $T_i$, $\ent{T_i|\setX \setminus S}=0$.  Then applying \Szero we re-order the parts so that $S, T_1, \ldots, T_k$ come first.  This yields,
\[
    \Icupe{S, T_1, \ldots, T_k}{y} \leq \Icupe{S, T_1, \ldots, T_k, \setX \setminus S}{y} \; .
\]

Applying \Mone inequality condition, adding the predictor $\setX \setminus S$ can only increase the union information.  Therefore we prove eq.~\eqref{eq:IbPunion1}, which proves eq.~\eqref{eq:IbBtobeproven}, that $\IbP(\setX~:~y )~=~\IbPTwo(\setX~:~y)$.
\end{proof}

\clearpage
\subsection{Bounds on $\psi(X_1, \ldots, X_n : y)$}
\label{appendix:psibounds}

{\lem{Given \Mone, \SR and the predictors $X_1, \ldots, X_n$ are independent, i.e., $\ent{X} = \sum_{i=1}^n \ent{X_i}$, then,
\[
    \psi(X_1, \ldots, X_n :y) \leq \min_{i \in \{1, \ldots, n\}} \DKL{ \Prob{X \middle|y} }{ \Prob{X_i} \Prob{X_{\sim i} \middle| y} } \; .
\] \label{lem:psiupper} }} \begin{proof}

Applying \Mone inequality condition, we have $\Icape{A,B}{y} \leq \min \left[ \info{A}{y}, \info{B}{y} \right]$.  Via the inclusion-exclusion rule, this entails $\Icupe{A,B}{y} \geq \max\left[ \info{A}{y}, \info{B}{y} \right]$, and we use this to upperbound $\psi(X_1, \ldots, X_n : y)$.  The random variable $A \not=\emptyset$, $B \equiv \setX \setminus A$, and $AB \equiv X$.

\begin{align*}
    \psi(X_1, \ldots, X_n : y ) &= \info{X}{y} - \max_{A \subset \setX} \Icupe{A,B}{y} \\
    &\leq \info{X}{y} - \max_{A \subset \setX} \max\left[ \info{A}{y}, \info{B}{y} \right] \\
    & \textnormal{By symmetry of complementary bipartitions, every $B$ will be an $A$ at some} \\
    & \textnormal{point.  So we can drop the $B$ term.} \\
    &= \info{X}{y} - \max_{A \subset \setX} \info{A}{y} \; .
\end{align*}

For two parts $A$ and $A^\prime$ such that $\ent{ A | A^\prime } = 0$, $\info{A}{y} \leq \info{A^\prime}{y}$.\footnote{$\info{A}{y} \leq \info{A^\prime}{y}$ because  $\info{A^\prime}{y} = \info{A}{y} + \info{A^\prime}{y|A}$.}  Therefore there will always be a maximizing subset of $\setX$ of size $n-1$.
\begin{align*}
   \psi(X_1, \ldots, X_n : y ) &\leq \info{X}{y} - \max_{\substack{A \subset \setX \\ |A| = n-1}} \info{A}{y} \\
   &= \info{X}{y} - \max_{i \in \{1, \ldots, n\}} \info{X_{\sim i}}{y} \\
   &= \min_{i \in \{1, \ldots, n\}} \info{X}{y} - \info{X_{\sim i}}{y} \\\ 
   &= \min_{i \in \{1, \ldots, n\}} \info{X_i}{y\middle|X_{\sim i}} \\
   &= \min_{i \in \{1, \ldots, n\}} \DKL{ \Prob{X\middle|y} }{ \Prob{X_i\middle|X_{\sim i}} \Prob{ X_{\sim i} \middle| y} } \; .
\end{align*}

Now applying that the predictors $\setX$ are independent, $\Prob{x_i\middle|x_{\sim i}}= \Prob{x_i}$.  This leaves,
\[
 \psi(X_1, \ldots, X_n :y) \leq \min_{i \in \{1, \ldots, n\}} \DKL{ \Prob{X \middle|y} }{ \Prob{X_i} \Prob{X_{\sim i} \middle| y} } \; .
\]
\end{proof}

\clearpage

{\lem{Given \GP, \SR and predictors $X_1, \ldots, X_n$ are independent, i.e., $\ent{X} = \sum_{i=1}^n \ent{X_i}$, then,
\begin{equation*}
\begin{split}
    \psi(X_1, \ldots, X_n : y) &\geq \min_{A \subset \setX} \info{A}{B\middle|y} \\
    &= \min_{A \subset \setX} \DKL{ \Prob{X|y}}{ \Prob{A|y} \Prob{B|y} } \; .
\end{split}
\end{equation*} \label{lem:psilower} }} \begin{proof}

First, from the definition of $\Icup$, $\Icupe{A,B}{y} = \info{A}{y} + \info{B}{y} - \Icape{A,B}{y}$.  Then applying \GP, we have $\Icupe{A,B}{y} \leq \info{A}{y} + \info{B}{y}$.  We use this to lowerbound $\psi(X_1, \ldots, X_n:y)$.  The random variable $A \not=\emptyset$, $B \equiv \setX \setminus A$, and $AB \equiv X$.

\begin{align*}
    \psi(X_1, \ldots, X_n:y) &= \info{X}{y} - \max_{A \subset \setX} \Icupe{A,B}{y} \\
    &\geq \info{X}{y} - \max_{A \subset \setX} \left[ \info{A}{y} + \info{B}{y} \right] \\
    &= \min_{A \subset \setX} \info{AB}{y} - \info{A}{y} - \info{B}{y} \\
    &= \min_{A \subset \setX} \info{A}{y|B} - \info{A}{y} \\
    &= \min_{A \subset \setX} \DKL{ \Prob{AB\middle|y} }{ \Prob{B|y} \Prob{A|B} } - \DKL{ \Prob{A\middle|y} }{ \Prob{A} } \\
    &= \min_{A \subset \setX} \sum_{a,b} \Prob{ab|y} \log \frac{\Prob{ab|y}}{\Prob{b|y} \Prob{a|b}} + \sum_{a} \Prob{a|y} \log \frac{ \Prob{a} }{ \Prob{a|y} } \; .
\end{align*}

We now add $\sum_{b} \Prob{b\middle|ay}$ in front of the right-most $\sum_{a}$.  We can do this because $\sum_{b} \Prob{b\middle|ay} = 1.0$.  Then yields,

\begin{align*}
    \psi(X_1, \ldots, X_n:y) &\geq \min_{A \subset \setX} \sum_{a,b} \Prob{ab|y} \log \frac{\Prob{ab|y}}{\Prob{b|y} \Prob{a|b}} + \Prob{b\middle|ay} \Prob{a|y} \log \frac{ \Prob{a} }{ \Prob{a|y} } \\
    &= \min_{A \subset \setX} \sum_{a,b} \Prob{ab|y} \left[ \log \frac{\Prob{ab|y}}{\Prob{b|y} \Prob{a|b}} + \log \frac{ \Prob{a} }{ \Prob{a|y} } \right] \\
    &= \min_{A \subset \setX} \sum_{a,b} \Prob{ab\middle|y} \log \frac{\Prob{ab|y} \Prob{a} }{ \Prob{a|y} \Prob{b|y} \Prob{a|b} } \; .
\end{align*}

Now applying that the predictors $\setX$ are independent, $\Prob{a|b}=\Prob{a}$; thus we can cancel $\Prob{a}$ for $\Prob{a|b}$.  This yields,
\begin{align*}
\psi(X_1, \ldots, X_n:y) &\geq \min_{A \subset \setX} \sum_{a,b} \Prob{ab\middle|y} \log \frac{\Prob{ab|y} }{ \Prob{a|y} \Prob{b|y} } \\
    &= \min_{A \subset \setX} \DKL{ \Prob{X|y}}{ \Prob{A|y} \Prob{B|y} } \; .
\end{align*}

\end{proof}

\clearpage

\subsection{Bounds on $\bpsi(X_1, \ldots, X_n : Y)$}
\label{appendix:bpsibounds}

{\lem{Given \Mone, \SR and the predictors $X_1, \ldots, X_n$ are independent, i.e., $\ent{X} = \sum_{i=1}^n \ent{X_i}$, then,
\[
    \bpsi(X_1, \ldots, X_n : Y) \leq \min_{i \in \{1, \ldots, n\}} \DKL{ \Prob{X, Y} }{ \Prob{X_{\sim i}, Y} \Prob{X_i} } \; .
\]}} \begin{proof}

First, using the same reasoning in Lemma \ref{lem:psiupper}, we have,
\begin{align*}
    \bpsi( \setX : Y ) &\leq \info{X}{Y} - \max_{i \in \{1, \ldots, n\} } \info{X_{\sim i}}{Y} \\
    &= \min_{i \in \{1, \ldots, n\} } \info{X}{Y} - \info{X_{\sim i}}{Y} \\
    &= \min_{i \in \{1, \ldots, n\} } \info{X_i}{Y\middle| X_{\sim i} } \\
    &= \min_{i \in \{1, \ldots, n\} } \DKL{ \Prob{X,Y} }{ \Prob{X_i \middle| X_{\sim i}} \Prob{ X_{\sim i}, Y} } \; .
\end{align*}

Now applying that the predictors $\setX$ are independent, $\Prob{X_i\middle|X_{\sim i}}= \Prob{X_i}$.  This yields,
\[
    \bpsi( \setX : Y ) \leq \min_{i \in \{1, \ldots, n\} } \DKL{ \Prob{X,Y} }{ \Prob{ X_{\sim i}, Y} \Prob{X_i} } \; .
\]
\end{proof}

{\lem{Given \GP, \SR and predictors $X_1, \ldots, X_n$ are independent, i.e., $\ent{X} = \sum_{i=1}^n \ent{X_i}$, then,
\begin{equation*}
    \bpsi(X_1, \ldots, X_n : Y) \geq \min_{A \subset \setX} \info{A}{B\middle|Y} \; .
\end{equation*} }} \begin{proof}

First, using the same reasoning in Lemma \ref{lem:psilower}, we have,

\begin{align*}
    \bpsi(X_1, \ldots, X_n:Y) &\geq \info{X}{Y} - \max_{A \subset \setX} \left[ \info{A}{Y} + \info{B}{Y} \right] \\
    &= \min_{A \subset \setX} \info{AB}{Y} - \info{A}{Y} - \info{B}{Y} \\
    &= \min_{A \subset \setX} \info{A}{B\middle| Y} - \info{A}{B} \; .
\end{align*}

Now applying that the predictors $\setX$ are independent, $\info{A}{B}=0$.  This yields,
\begin{equation*}
    \bpsi(X_1, \ldots, X_n:Y) \geq \min_{A \subset \setX} \info{A}{B\middle|Y} \; .
\end{equation*}
\end{proof}

\clearpage

\section{Definition of Intrinsic $\ei$ a.k.a. ``Perturbing the Wires''}
\label{appendix:perturbingwires}
State-dependent $\ei$ across a partition, $\ei\!\left( X \to y / \P\right)$, is defined by eq.~\eqref{eq:statedepei}.

\begin{eqnarray}
	\label{eq:statedepei}
\nonumber	\ei \! \left( X \to y / \P \right) &\equiv& \DKL{ \Prob{X \to y} }{ \prod_{i=1}^m \Prob{ X^\P_i \to y^\P_i } } \\ 
\label{eq:statedepei2}
	&=& 	 \DKL{ \Prob{X|y} }{ \prod_{i=1}^m \Probstar{ X^\P_i \middle| y^\P_i } } \; .
\end{eqnarray}

Balduzzi/Tononi \cite{balduzzi-tononi-08} define the probability distribution describing the intrinsic information from the whole system $X$ to state $y$ as,
\[
	\Prob{ X \to y} = \Prob{ X | y} = \left\{ \Prob{ x | y } : \forall x \in X \right\} \; .
\]

They then define probability distribution describing the intrinsic information from a part $X^\P_i$ to a state $y^\P_i$ as,
\[
	\Probstar{ X^\P_i \to y^\P_i} \equiv \Probstar{ X^\P_i \middle| y^\P_i } = \left\{ \Probstar{ x^\P_i \middle| y^\P_i } : \forall x^\P_i \in X^\P_i \right\} \; .
\]

First we define the fundamental property of the $\Pr^*$ distribution.\footnote{It's worth noting that $\Probstar{X|y} \not= \Prob{X|y}$.}  Given a state $x^\P_i$, the probability of a state $y^\P_i$ is computed by probability each node in the state $y^\P_i$ independently reaches the state specified by $y^\P_i$,

\begin{equation}
    \Probstar{ y^\P_i \middle| x^\P_i } \equiv \displaystyle \prod_{j=1}^{|\P_i|}\Prob{ y^\P_{i,j} \middle| x^\P_i } \; .
\label{eq:product}
\end{equation}

Then we define the join distribution relative to eq.~\eqref{eq:product}:
\[
    \Probstar{x^\P_i, y^\P_i} = \Probstar{x^\P_i} \Probstar{y^\P_i  \middle| x^\P_i } = \Probstar{x^\P_i} {\displaystyle \prod_{j=1}^{|\P_i|} } \Prob{ y^\P_{i,j} \middle| x^\P_i } \; .
\]

Then applying assumption \textbf{(B)}, $X$ follows a discrete uniform distribution, so $\Probstar{x^\P_i}~\equiv~\Prob{x^\P_i} = 1 / |X^\P_i|$.  This gives us the complete definition of $\Probstar{x^\P_i, y^\P_i}$,
\begin{equation}
    \Probstar{x^\P_i, y^\P_i} = \Prob{x^\P_i} \prod_{j=1}^{|\P_i|}\Prob{ y^\P_{i,j} \middle| x^\P_i } \; .
\label{eq:pstarjoint}
\end{equation}

With the joint $\Pr^*$ distribution defined, we can compute anything we want---such as the expressions for $\Probstar{y^P_i}$ and $\Probstar{x^\P_i \middle| y^\P_i}$---by summing over eq.~\eqref{eq:pstarjoint},
\begin{align}
    \Probstar{y^\P_i} &= \sum_{x^\P_i \in X^\P_i} \Probstar{x^\P_i, y^\P_i} \\
	\Probstar{ x^\P_i \middle| y^\P_i } &= \frac{\Probstar{x^\P_i, y^\P_i}}{\Probstar{y^\P_i}} \; .
\end{align}

\clearpage

\section{Setting $t=1$ Without Loss of Generality}
\label{appendix:singletime}

Given $t$ stationary surjective functions that may be different or the same, denoted $f_1 \cdots f_t$, we define the state of system at time $t$, denoted $X_t$, as the application of the $t$ functions to the state of the system at time $0$, denoted $X$,

\[
    X_t = f_t\left( f_{t-1}\left( \cdots f_{2}\left( f_1\left(X\right) \right) \cdots \right) \right) \; .
\]

We instantiate an empty ``dictionary function'' $g\left(\bullet\right)$.  Then for every $x_0 \in X_0$ we assign,
\[
	g\left( x\right) \equiv  f_t\left( f_{t-1}\left( \cdots f_{2}\left( f_1\left(x\right) \right) \cdots \right) \right) \; .
\]

At the end of this process we have a function $g$ that accomplishes any chain of stationary functions $f_1 \cdots f_t$ in a single step for the entire domain $X$.  So instead of studying the transformation, 
\[
X \overset{f_1 \cdots f_t}{\longrightarrow} X_t \; ,
\]
we can equivalently study the transformation,
\[
X \overset{g}{\longrightarrow} Y \; .
\]

Here's an example using mechanism $f_1 = f_2 = f_3 = f_4 = \textnormal{AND-GET}$.

\begin{table}[h!]
  	\centering
	\begin{tabular}{  c c c c c c c c c c c } \toprule
		time=0 & \ & $t=1$& \ & $t=2$ & \ & $t=3$ & \ & $t=4$ \\
		\midrule
		\bin{00} &$\to$& \bin{00} &$\to$& \bin{00} &$\to$& \bin{00} &$\to$& \bin{00} \\
		\bin{01} &$\to$& \bin{00} &$\to$& \bin{00} &$\to$& \bin{00} &$\to$& \bin{00} \\
		\bin{10} &$\to$& \bin{01} &$\to$& \bin{00} &$\to$& \bin{00} &$\to$& \bin{00} \\
		\bin{11} &$\to$& \bin{11} &$\to$& \bin{11} &$\to$& \bin{10} &$\to$& \bin{00} \\
		\bottomrule
		\addlinespace
		$g\left( \bullet \right)$ & \ & AND-GET & \ & AND-AND & \ & AND-ZERO & \ & ZERO-ZERO \\
	\end{tabular}
	\caption{Applying the update rule ``AND-GET'', over four timesteps.}
\end{table}


\section{The Appropriate Distribution on $X$ is Ambiguous}
\label{appendix:Xdistribution}
A system's ``mechanism'' is defined by the probability distribution $\Prob{Y|X}$.  And we are asking that given a state $Y=y$, how clearly are the possible states of $X$ specified---i.e., Given the mechanism $\Prob{Y|X}$, how different are the distributions $\Prob{X}$ and $\Prob{X|y}$?  To compute $\Prob{X|y}$ from $\Prob{Y|X}$, we must define a distribution $\Prob{X}$.  There are several choices for $\Prob{X}$.  These are same of the prominent ones:

\begin{description}
    \item[Empirical:] Make $X$ follow the distribution actually recorded from the system.

    \item[Discrete uniform:] Every state $x \in X$ has $\Prob{x} = \frac{1}{|X|}$ where $|X|$ is the number of distinct states of r.v. $X$.
    
    \item[Capacity:]  Regardless of state $y \in Y$, the $X$ distribution is,
\[
    X \sim \argmax_{\Pr(X^\prime)} \info{X^\prime}{Y}
\]
\end{description}

Each of these distributions have been used for causal measures\cite{ay-06,nyberg09, janzing12}.  And for each of these candidate distributions on $X$, there exist (causal) questions for which it is the best/most appropriate choice.  Therefore, merely saying we want a ``causal measure'' for conscious experience does not rule any of them out.  Conceptually, it makes sense to preclude the empirical distribution as it does not take into account counterfactuals.  But what about the discrete-uniform versus the capacity distribution?  What reason is there to prefer one over the other?  Ideally this would be answered by returning to the original thought experiments for consciousness.

\end{document}